\documentclass[10pt,twocolumn,oneside,final]{IEEEtran}

\IEEEoverridecommandlockouts
\usepackage{cite}
\usepackage{graphicx}
\usepackage{amsmath}
\usepackage{times}
\usepackage{latexsym}
\usepackage{bm}
\usepackage{amssymb}
\usepackage[center]{caption2}
\usepackage{array}
\usepackage{fancyhdr}
\usepackage{cite,graphicx,amsmath,amssymb}
\usepackage{citesort}
\usepackage{psfrag}
\usepackage{multirow}
\usepackage{amsthm}

\ifCLASSINFOpdf

\else

\fi
\usepackage[dvipsnames,usenames]{color}
\usepackage[usenames,dvipsnames]{color}

\newcommand{\bl}[1]{\color{black}#1}

\newtheorem{theorem}{Theorem}

\newtheorem{corollary}{Corollary}

\newcommand{\tr}{\text{tr}}

\makeatother

\begin{document}
\title{Data-Aided Secure Massive MIMO Transmission under the Pilot
Contamination Attack}

\author{\IEEEauthorblockN{Yongpeng Wu, Chao-Kai Wen, Wen Chen, Shi Jin, Robert Schober,
and  Giuseppe Caire}

\thanks{This paper was presented in part at IEEE ICC 2018.}

\thanks{The work of Y. Wu is supported in part by the National Science Foundation
(NSFC) under Grant 61701301, Young Elite Scientist Sponsorship Program
by CAST, and the open research project of State Key Laboratory of Integrated
Services Networks (Xidian University) under Grant ISN20-03. The work of C.-K. Wen
was supported in part by the Ministry of Science and Technology of Taiwan under grants MOST 107-2221-E-110-026.
The work of
W. Chen is supported in part by NSFC under Grant 61671294, STCSM
Project under Grant 16JC1402900 and 17510740700, National Science
and Technology Major Project under Grant 2018ZX03001009--002.
The work of S. Jin was supported in part by the National Science Foundation (NSFC)
for Distinguished Young Scholars of China with Grant 61625106.}

\thanks{Y. Wu is with the Department of Electronic Engineering, Shanghai Jiao Tong University,
Minhang 200240, China (e-mail: yongpeng.wu@sjtu.edu.cn). Y. Wu is also with State Key Laboratory of Integrated Services Networks,
Xidian University, Xi’an, China.}

\thanks{C. K. Wen is with the Institute of Communications Engineering, National Sun Yat-sen University, Kaohsiung 804,
Taiwan (Email: chaokai.wen@mail.nsysu.edu.tw).}

\thanks{W. Chen is with Shanghai Institute of Advanced Communications and Data Sciences, the Department of Electronic
Engineering, Shanghai Jiao Tong University, Minhang 200240, China (E-mail: wenchen@sjtu.edu.cn).}

\thanks{S. Jin is with the National Mobile Communications Research Laboratory,
Southeast University, Nanjing, 210096, P. R. China. (Emails: jinshi@seu.edu.cn).}

\thanks{R. Schober is with Institute for Digital Communications, Universit\"{a}t Erlangen-N\"{u}rnberg,
Cauerstrasse 7, D-91058 Erlangen, Germany (Email: schober@lnt.de).}

\thanks{G. Caire is with Institute for Telecommunication Systems, Technical University Berlin, Einsteinufer 25,
10587 Berlin, Germany (Email: caire@tu-berlin.de). }

}

\maketitle


\begin{abstract}
In this paper, we study the design of secure communication for time division duplex
multi-cell multi-user massive multiple-input multiple-output (MIMO) systems with active
eavesdropping. We assume that the eavesdropper actively attacks the uplink
pilot transmission and the uplink data transmission before eavesdropping
the downlink data  transmission  of the users.
We exploit both the received pilots and the received data signals for uplink channel
estimation.  We show analytically
that when the number of transmit antennas and the length of the data vector both tend to infinity,
the signals of the desired user and the eavesdropper lie in
different eigenspaces of the received signal matrix at the base station
provided that their signal powers are different.
This finding reveals that  decreasing (instead of increasing)
the desired user's signal power might be an effective approach
to combat a strong active attack
from an eavesdropper. Inspired by this observation, we propose a data-aided
secure downlink transmission scheme and derive an asymptotic achievable secrecy
sum-rate expression for the proposed design.
For the special case of a single-cell single-user system with
independent and identically distributed fading, the obtained
expression reveals that the secrecy rate scales logarithmically
with the number of transmit antennas. This is the same scaling
law as for the achievable rate of  a single-user massive MIMO system in the absence of eavesdroppers.
Numerical results indicate that the proposed scheme
achieves significant secrecy rate gains compared  to
alternative approaches based on
matched filter precoding with artificial noise
generation and null space transmission.
\end{abstract}

\section{Introduction}
Wireless networks are widely used in civilian and military applications
and have become an indispensable part of our daily lifes. Therefore, secure communication
is a critical issue for future wireless networks.  {\bl As a complement
to the conventional cryptographic techniques, new  approaches to secure communication based on information theoretical
concepts, such as the secrecy capacity of the propagation channel, have been
developed and are collectively referred to as physical layer security \cite{Wyner1975BSTJ,Oggier2011TIT,Wu2012TVT,Wu2017TCom,Wu2018JSAC,Li2018CM}.}

Multiple-input multiple-output (MIMO)
technology has been shown to be a promising means for providing multiplexing gains and diversity gains,
leading to improved performance \cite{Wu2012TWC,Wu2013TCOM,Wu2014TSP,Chen2014CM,Chen2016CM,Wu2018P}. In particular, massive MIMO technology, which
utilizes a very large number of antennas at the BS
and simple signal processing to provide services for a comparatively small (compared to the number of antennas)
number of active  mobile users, is a promising approach for efficient transmission of massive amounts of information
and is regarded as {\bl a key technology} in 5G \cite{Andrews2014JSAC}.
Pilot contamination is a major impairment in massive MIMO systems \cite{Marzetta2010TWC} and many approaches
have been proposed to solve this problem \cite{Wen2015ICC,Muller2014JSTSP,Haghighatshoar2016TWC,Ni2017TWC}. For example, the authors of
\cite{Ni2017TWC} proposed an approach based on Chu sequences,
which efficiently reduces the effect of pilot contamination.

Most studies on physical layer security in massive MIMO systems
assume that the eavesdropper is passive and does not attack the communication
process of the system \cite{Zhu2014TWC,Chen2015TWC,Chen2016TFS,ZhuJ2017TWC,Nguyen2018JSAC}.
However, a smart eavesdropper
can perform a pilot contamination attack to impair
the channel estimation process at the base station \cite{Wu2016TIT}.
Due to the channel hardening effect caused by large antenna arrays,
it is difficult to exploit the statistical fluctuations of
fading channels to safeguard the transmission.
Then, the beamforming direction misled
by the pilot contamination attack can significantly enhance the performance
of the eavesdropper. This results in a serious secrecy  threat
in time division duplex (TDD)-based massive MIMO systems \cite{Wu2016TIT}.

Prior works on the pilot contamination attack have studied mechanisms for
enhancing the eavesdropper's performance \cite{Zhou2012TWC,Xu2016WCL,Nguyen2017TCOM,Huang2018TWC}.
Other works propose various approaches for detecting the pilot
contamination attack \cite{Kapetanovic2013PIMRC,Kapetanovic2015CM,Xiong2015TIFS,Tugnait2015WCL,Xiong2016TIFS}.
Although these schemes
can detect a pilot contamination attack with high probability,
 \cite{Kapetanovic2013PIMRC,Kapetanovic2015CM,Xiong2015TIFS,Tugnait2015WCL,Xiong2016TIFS}
do not provide an effective
transmission scheme in the presence of a pilot contamination attack.
For secure communication under a pilot contamination attack,
the authors of \cite{Im2015TWC} propose a secret key agreement
protocol for single-cell multi-user massive MIMO systems.
An estimator for the base station (BS)
is designed to evaluate the resulting information leakage.
Then, the BS and the desired users perform secure communication by adjusting the length
of the secret key based on the estimated information leakage.
Other works have studied how to combat the pilot contamination attack.
The authors of \cite{Basciftci2017} investigate
the pilot contamination attack problem for single-cell multi-user massive MIMO systems
over independent and identically distributed (i.i.d.) fading channels.
The eavesdropper is assumed to only know the pilot signal set whose size scales polynomially
with the number of transmit antennas. For each transmission, the desired users randomly select certain pilot signals
from this set, which are unknown to the eavesdropper.
In this case, it is proved that the impact of the pilot contamination
attack can be eliminated as the number of transmit antennas goes to infinity.
Under the same pilot allocation protocol, the authors of \cite{Wang2018JSAC} and \cite{Do2018TIFS}
respectively propose a random channel training scheme and a jamming-resistant scheme employing an unused
pilot sequence to combat the  pilot contamination
attack and to maintain secure communication. Moreover,
by exploiting an additional random sequence, which is transmitted by the
legitimate users but is unknown to the eavesdropper, an effective blind channel
estimation method and a secure beamforming scheme are developed
to realize  reliable transmission in \cite{Tugnait2018TCom}.

However, all the above-mentioned methods rely on a key assumption:
some form of an additional pilot signal protocol which is unknown to the eavesdropper
is needed to combat the pilot contamination attack.
For the more pessimistic case where the
eavesdropper knows the desired users' exact pilot signal structure for each transmission\footnote{We note that this case
constitutes a worst-case scenario. Hence, if secure communication
can be achieved for this worst case, then secure communication can also be achieved for more optimistic settings as considered
in \cite{Wang2018JSAC,Do2018TIFS,Tugnait2018TCom}.},
the secrecy threat caused by
the pilot contamination attack in multi-cell multi-user massive MIMO systems over correlated fading channels is analyzed in \cite{Wu2016TIT}.
Based on this analysis, three transmission strategies for combating the pilot contamination attack
are proposed. Nevertheless, the designs in \cite{Wu2016TIT} are not able to guarantee a high (or not even a non-zero) secrecy rate for
weakly correlated or i.i.d. fading channels when the power of the eavesdropper pilot signal
is much larger than that of the users' pilot signals.

In this paper, we investigate secure transmission for
TDD multi-cell multi-user massive MIMO systems
impaired by  general correlated fading and a pilot contamination attack.
We assume the considered system performs first uplink training followed by uplink and downlink data transmission phases.  The eavesdropper jams the uplink training phase
and the uplink data transmission phase and then eavesdrops the downlink data transmission\footnote{\bl
We note that the pilot contamination attack presents a security threat \textit{only} for the downlink.
In fact, in the uplink, if an eavesdropper transmits a strong pilot signal in the uplink together with
the legitimate  users, it can at most jam the reception of some users, and therefore disrupt coherent detection,
but it will not be able to improve its ability to eavesdrop
the uplink traffic. Since this paper studies the pilot contamination attack, we focus on downlink data transmission.
Secure uplink transmission is also relevant and is an interesting topic for future work (not dedicated to the pilot contamination attack)
but is outside the scope of the present paper.} We utilize the data transmitted in the uplink to aid
the channel estimation at the BS. Then, based on the estimated channels,
the BS designs precoders for downlink transmission.

This paper makes the following key contributions:

\begin{enumerate}

\item We prove that when the number of transmit antennas and the amount of
transmitted data both approach infinity, the desired users' and the eavesdropper's signals lie in
different eigenspaces of the uplink received signal matrix
provided that their pilot signal powers are different.  Our results reveal
that increasing the power gap between the desired users' and the eavesdropper's signals
is beneficial for separating the desired users and the eavesdropper.
This implies that when facing an active attack, decreasing (instead of increasing)
the desired users' signal power could be an effective approach for enabling secrete
 communication.

\item Inspired by this observation, we propose a joint uplink and downlink data-aided transmission
scheme to combat strong\footnote{We refer to a pilot contamination attack  as a strong pilot contamination attack  if
the pilot signal power of the eavesdropper is much larger than that of the users.} active attacks from an eavesdropper.
Then, we derive an asymptotic expression for the corresponding achievable secrecy
sum-rate. The derived
expression indicates that the impact of an active attack on  uplink
transmission can be effectively eliminated by the proposed design.

\item We specialize the asymptotic achievable secrecy
sum-rate expression to the case of i.i.d. fading
channels. Particularly, for the classical
MIMO eavesdropper wiretap model,
the derived
expression indicates that the secrecy rate exhibits a logarithmic growth
with the number of transmit antennas.
This is the same growth rate as that  of the achievable rate
 of a typical point-to-point massive MIMO system without eavesdropping.

\end{enumerate}

The remainder of this paper is organized as follows. In Section II,
the basic system model is introduced and the uplink channel estimation is investigated.
In Section III, the proposed secure downlink transmission scheme is presented
and an asymptotic expression for the secrecy rate of the proposed design is derived. Section IV
discusses the special case of i.i.d. fading in detail.
Numerical results are provided in Section V,
and conclusions are drawn in Section VI.

\emph{Notation:}  Vectors are denoted by lower-case bold-face letters;
matrices are denoted by upper-case bold-face letters. Superscripts $(\cdot)^{T}$, $(\cdot)^{*}$, and $(\cdot)^{H}$
stand for the matrix transpose, conjugate, and conjugate-transpose operations, respectively. We use  ${\tr}({\bf{A}})$ and ${\bf{A}}^{-1}$
to denote the trace and the
inverse of matrix $\bf{A}$, respectively.
 ${\rm{diag}}\left\{\bf{b}\right\}$ denotes a diagonal matrix
with the elements of vector $\bf{b}$ on its main diagonal.
${\rm{Diag}}\left\{\bf{B}\right\}$  denotes a diagonal matrix containing
the diagonal elements of matrix $\mathbf{B}$ on the main diagonal.
The $M \times M$ identity matrix is denoted
by ${\bf{I}}_M$, and the $M \times N$ all-zero matrix and the $N \times 1$ all-zero vector are denoted by $\bf{0}$.
The fields of complex and real numbers are denoted
by $\mathbb{C}$ and $\mathbb{R}$, respectively. $E\left[\cdot\right]$ denotes statistical
expectation. $[\mathbf{A}]_{mn}$ denotes the element in the
$m$th row and $n$th column of matrix $\mathbf{A}$. $[\mathbf{a}]_{m}$ denotes the $m$th entry
of vector $\mathbf{a}$. $\otimes$ denotes the Kronecker product.
$\mathbf{x} \sim \mathcal{CN} \left( {\mathbf{0},{{\bf{R}}_N}} \right)$
denotes a circularly symmetric complex vector
$\mathbf{x} \in {\mathbb{C}^{N \times 1}}$  with zero mean and covariance matrix
${{\bf{R}}_N}$.
${\rm var} (a) $ denotes  the variance of  random variable $a$.
${\left[ x \right]^ + }$ stands for $\max \left\{ {0,x} \right\}$.
$a \gg b$ means that $a$ is much larger than $b$.  $f \in {o}(x)$
means that $f/x\rightarrow 0$.

\section{Uplink Transmission} \label{sec:multi}
Throughout the paper, we adopt the following transmission protocol.
We assume the uplink transmission phase, comprising uplink
training and  uplink data transmission,
is followed by a downlink data transmission phase.

We assume the main objective of the
eavesdropper is to eavesdrop the downlink data. Nevertheless,
the eavesdropper also attacks  the
uplink transmission phase to impair the channel estimation at the BS.
The resulting mismatched channel estimation will increase the information leakage in the subsequent downlink transmission.
In the downlink transmission phase, the eavesdropper does not attack
but focuses on eavesdropping the data.

\begin{figure*}[!ht]
\centering
\includegraphics[width=0.8\textwidth]{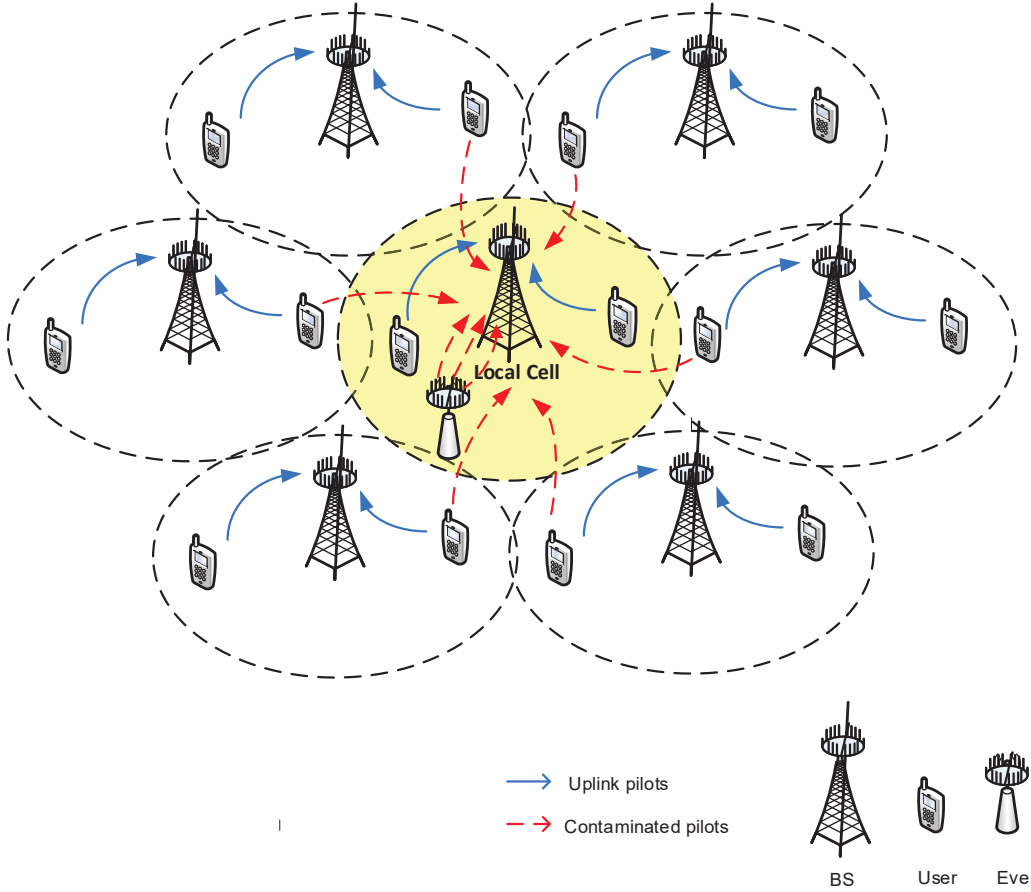}
\caption {\space\space Model of considered multi-cell massive MIMO system.}
\label{system model}
\end{figure*}

We study a multi-cell multi-user MIMO system with $L + 1$ cells, cf. Figure \ref{system model}.
We assume an $N_t$-antenna BS and $K$ single-antenna users are present in
each cell. The cells are index by $l = \left(0,\ldots,L\right)$,
where cell $l = 0$ is the cell of interest.
We assume an $N_e$-antenna
active eavesdropper\footnote{\bl An $N_e$-antenna eavesdropper is equivalent to
$N_e$ cooperating single-antenna eavesdroppers.}
is located in the cell of interest
and attempts to eavesdrop the data intended for all users in the cell.
The eavesdropper sends pilot signals and artificial noise
to interfere channel estimation and uplink data transmission\footnote{We note that if the eavesdropper only attacks the channel
estimation phase and remains silent during the uplink data transmission, then the impact of this attack can be easily eliminated
with the joint channel estimation and data detection scheme in \cite{Wen2015ICC}. Therefore, a smart eavesdropper will attack the entire uplink transmission.},
respectively. Let $T$ and $\tau$ denote the coherence time of the channel and the length of
the pilot signal, respectively.  Then, for uplink transmission, the received pilot signal matrix
$\mathbf{Y}_p^{m} \in {\mathbb{C}^{N_t \times \tau}} $ and the received data signal matrix $\mathbf{Y}_d^{m} \in {\mathbb{C}^{N_t \times (T - \tau)}}$
at the BS in cell $m$ are given by
\begin{align}\label{eq:Yp}
{\mathbf{Y}}_{p}^{m} & =\sqrt {{P_{0}}} \sum\limits_{k = 1}^K { {\bf{h}}_{0k}^m{\boldsymbol{\omega }}_{k}^T}  +  \sum\limits_{l = 1}^L {\sum\limits_{k = 1}^K { \sqrt {{P_{l}}} {\bf{h}}_{lk}^m{\boldsymbol{\omega }}_{k}^T} } \nonumber \\
& + \sqrt {\frac{{{P_e}}}{{K{N_e}}}} {\bf{{H}}}_{e}^m  {\mathbf{W}_e} + {\mathbf{N}_p^m},
\end{align}
\begin{align}\label{eq:Yd}
{\mathbf{Y}}_{d}^{m}   & = \sqrt {{P_{0}}} \sum\limits_{k = 1}^K { {\bf{h}}_{0k}^m{\mathbf{d}}_{0k}^T}  +  \sum\limits_{l = 1}^L {\sum\limits_{k = 1}^K { \sqrt {{P_{l}}} {\bf{h}}_{lk}^m {\mathbf{d}}_{lk}^T} } \nonumber \\
 & + \sqrt {\frac{{{P_e}}}{{N_e}}} {\bf{{H}}}_{e}^m \mathbf{A}  + {\mathbf{N}_d^m},
\end{align}
where $P_{0}$,  ${{\boldsymbol{\omega}}_{k}} \in \mathbb{C} {^{\tau  \times 1}}$,
and ${\mathbf{d}}_{0k} \sim \mathcal{CN} \left( {\mathbf{0},{{\bf{I}}_{T - \tau}}} \right)$ denote the average transmit power, the pilot sequence,
and the uplink transmission data  of the $k$th user in the cell of interest, respectively.
For simplicity of notation, we assume that all users in a given cell use the same transmit power \cite{Zhu2014TWC}.
Using  similar techniques as presented in this paper, our results can be easily extended to the case where
the users in a  cell have different transmit powers. We assume that the users in different cells have different powers.
It is assumed that the same $K$ orthogonal pilot sequences
are used in each cell where ${\boldsymbol{\omega }}_{k}^H{{\boldsymbol{\omega }}_{k}} = \tau$
and ${\boldsymbol{\omega }}_{k}^H{{\boldsymbol{\omega }}_{l}} = 0$, $k \neq l$.
${\mathbf{W}_e}$ is the  pilot attack signal of the eavesdropper.
$P_{l}$ and ${\mathbf{d}}_{lk}$ denote the average transmit power and
the uplink transmission data  of the $k$th user in the $l$th cell, respectively.
$\mathbf{h}_{lk}^p \sim \mathcal{CN} \left( {\mathbf{0}, {\mathbf{R}_{lk}^p}} \right)$
denotes the channel between the $k$th user in the $l$th cell and the BS in the $p$th cell,
where ${\mathbf{R} _{lk}^p}$ is the corresponding correlation matrix.
$\mathbf{H}_{e}^l$ and $P_e$ denote the channel between the eavesdropper and the BS in
the $l$th cell and the average transmit power of the eavesdropper, respectively.
$\mathbf{H}_{e}^l  \sim \mathcal{CN} \left(\mathbf{0}, {{\bf{R}}_{E,T}^l \otimes {\bf{R}}_{E,R}^l} \right)$
represents the channel between  the eavesdropper and the BS in the $l$th cell,
where ${\bf{R}}_{E,T}^l$ and  ${\bf{R}}_{E,R}^l$  are the corresponding
transmit and receive correlation matrices of the eavesdropper. $\mathbf{N}_p^m \in \mathbb{C} {^{N_t \times \tau}} $ and $ \mathbf{N}_d^m \in \mathbb{C} {^{N_t  \times (T - \tau)}} $ are noise matrices
whose columns are i.i.d. Gaussian distributed with $\mathcal{CN} \left( {\mathbf{0}, {N_0} {{\bf{I}}_{{N_t}}}} \right)$.

During the training phase, the eavesdropper attacks all users in the cell of interest.
In this paper, we adopt the worst-case assumption that for each transmission, the eavesdropper knows
the exact pilot sequence ${{\boldsymbol{\omega }}_{k}}$ of each user.
Therefore, it uses pilot attack sequences\footnote{If the eavesdropper is only interested in a particular user, then he can perform a pilot contamination attack specifically
for this user
as in \cite{Wu2016TIT}. However, this pilot contamination precoding will not influence the  proposed joint uplink and downlink transmission scheme
and the corresponding asymptotic performance analysis.} $ {\mathbf{W}_e} = \sum\nolimits_{k=1}^{K} {\mathbf{W}_k}$ \cite{Basciftci2017},
where $\mathbf{W}_k = \left[{{\boldsymbol{\omega}}_{k}} \cdots {{\boldsymbol{\omega}}_{k}} \right]^T \in \mathbb{C} {^{N_t  \times \tau}}$.
In the uplink data transmission phase, the eavesdropper generates an artificial noise matrix $\mathbf{A} \in \mathbb{C} {^{N_t \times T - \tau}}$,
whose elements follows an i.i.d. standard Gaussian distribution.

We define $\mathbf{Y}_0 = \left[{\mathbf{Y}}_{p}^{0} \quad  {\mathbf{Y}}_{d}^{0} \right]$ and the
eigenvalue decomposition $\frac{1}{{T{N_t}}}{\mathbf{Y}_0}{{\mathbf{Y}_0^H}} = \left[ {{{\bf{v}}_1},
\cdots ,{{\bf{v}}_{{N_t}}}} \right]{\bf{\Sigma }}{\left[ {{{\bf{v}}_1}, \cdots ,{{\bf{v}}_{{N_t}}}} \right.}$ \\
$ {\left. \right]^H}$, where the eigenvalues on the main diagonal of  matrix ${\bf{\Sigma }}$ are organized in ascending order.
For the following, we make the important assumption that
due to the difference in power\footnote{When the eavesdropper increases the
pilot contamination attack power $P_e$, the users in the cell of interest can intentionally decrease their pilot signal power
$P_0$ to achieve a significant power gap between $P_e$ and $P_0$ based on a power control mechanism. According to Appendix A, this is essential
to eliminate the impact of the active eavesdropper. If the eavesdropper does not eavesdrop the signals in the cell in which it is located but the signals
in one of the other cells,
then a power control mechanism for $P_e$ and $P_l$ is required to combat the pilot contamination attack.} between the eavesdroppers' pilots  and the users' pilots in the cell of interest
and the large path loss difference between the users in the cell of interest and the users in
the other cells \cite{Muller2014JSTSP},
 ${P_e} {\rm{tr}}\left( {{\bf{R}}_{E,T}^0} \right)$, ${P_0} {\rm{tr}} \left({\mathbf {R}_{0k}^0} \right)$, and $P_l {\rm{tr}} \left({\mathbf {R}_{lk}^l} \right)$
 have the relationship ${P_e} {\rm{tr}}\left( {{\bf{R}}_{E,T}^0} \right) \gg {P_0} {\rm{tr}} \left({\mathbf {R}_{0k}^0} \right) \gg P_l {\rm{tr}} \left({\mathbf {R}_{lk}^l} \right)$.
Define $ {\rm{tr}} \left( {{\bf{R}}_{E,T}^0} \right) {\bf{R}}_{E,R}^0= {{\bf{R}}_E}$ and
the eigenvalue decomposition ${{\bf{R}}_E} = \mathbf{U}_E  \boldsymbol{\Lambda}_e \mathbf{U}_E^H$,
where ${{\bf{\Lambda }}_{{e}}} = {\rm diag} \left( {{\Lambda _1} \cdots {\Lambda_{{N_{{e}}}}}} \right)$.
Let $M = (L + 1)K + {N_e}$ and  vector $\left( {{\theta _1}, \cdots ,{\theta _M}} \right)$ has the same
elements as  vector $\left({P_1} {\rm tr}\left(\mathbf{R}_{11}^0\right), \cdots ,{P_L} {\rm tr} \left(\mathbf{R}_{LK}^0 \right), {P_0}  {\rm tr} \left(\mathbf{R}_{01}^0 \right),
\cdots , {P_0} {\rm tr} \left( \mathbf{R}_{0K}^0 \right), \right.$ \\
 $\left.{P_e} \Lambda _1 , \cdots  {P_e} \Lambda_{N_e}\right)$
but with the elements organized in ascending order
such that index  $1 \le {i_1} \le {i_2} \cdots  \le {i_K} \le M$ satisfies ${\theta _{{i_k}}} = {P_0} {\rm tr} \left(\mathbf{R}_{0k}^0\right)$, $k = 1,2, \cdots ,K$.
Define $\mathbf{V}_{eq}^0 = \left[ {{{\bf{v}}_{{N_t} - M + {i_1}}},{{\bf{v}}_{{N_t} - M + {i_2}}}, \cdots, {{\bf{v}}_{{N_t} - M + {i_K}}}} \right]$,
${{\bf{H}}_0} = \left[ {{\bf{h}}_{01}^0, \cdots {\bf{h}}_{0K}^0} \right]$, and
${{\bf{H}}_I} = \left[ {{\bf{h}}_{11}^0, \cdots {\bf{h}}_{1K}^0, \cdots ,{\bf{h}}_{L1}^0, \cdots ,{\bf{h}}_{LK}^0} \right]$.
Then, we have the following theorem.

\begin{theorem}\label{theo:channel_estimation}
Let ${\mathbf{Z}_{0p}} =  \frac{1}{{\sqrt {T{N_t}} }} \left(\mathbf{V}_{eq}^{0}\right)^H \mathbf{Y}_{p}^{0} = \left[ {{{\bf{z}}_{0p,1}}, \cdots ,{{\bf{z}}_{0p,K}}} \right]$
and ${{\mathbf{H}}_{eq}^{0}} = \frac{1}{{\sqrt {T{N_t}} }}\left(\mathbf{V}_{eq}^{0}\right)^H{{\bf{H}}_0}  = \left[ { {{\bf{h}}_{eq,01}}, \cdots, } {{{\bf{h}}_{eq,0K}}} \right]$.
Then, when $T \rightarrow \infty $ and $N_t \rightarrow \infty$, the received signal $\mathbf{z} = {\rm vec}\left(\mathbf{Z}_{0p}\right)$
and the minimum mean square error (MMSE) estimate  ${{\bf{\widehat{h}}}_{eq,0k}}$ of ${{\bf{h}}_{eq,0k}}$
based on $\mathbf{z}$ are given by
\begin{align}
{\bf{z}}  & = \sqrt {{P_0}} \sum\limits_{t = 1}^K {\left( {{{\boldsymbol{\omega }}_t} \otimes {{\bf{I}}_K}} \right){{\bf{h}}_{eq,0t}}}  + {\bf{n}}, \label{eq:z} \\
  {{{\bf{\hat h}}}_{eq,0k}} &  {\rm{ = }}\sqrt {{P_0}} \mathbf{V}_{eq}^0 {\bf{R}}_{0k}^0 \left(\mathbf{V}_{eq}^0 \right)^H \nonumber \\
& \hspace{-1cm} \times {\left( \! {{N_0}{{\bf{I}}_K} + \tau {P_0} \mathbf{V}_{eq}^0 {\bf{R}}_{0k}^0 \left( \!\mathbf{V}_{eq}^0  \!\right)^H} \! \right)^{ - 1}}\left( \! {\sqrt {{P_0}} \tau {\mathbf{V}_{eq}^0} {\bf{h}}_{0k}^0 + {\bf{n}}_{eq} }  \!\right),  \label{eq:h_est}
\end{align}
 where
\begin{align}\label{eq:noise_theorem_1}
{\bf{n}} = \left[ \begin{array}{l}
{\left( {{\bf{V}}_{eq}^0} \right)^H}{{\bf{n}}_{p1}^{0}}\\
 \vdots \\
{\left( {{\bf{V}}_{eq}^0} \right)^H}{{\bf{n}}_{p\tau}^{0} }
\end{array} \right]
\end{align}
and ${\bf{n}}_{pt}^{0}$ in (\ref{eq:noise_theorem_1}) is the $t$th column of  $\mathbf{N}_p^0$. $ {\mathbf{{n}}_{eq}} = {{\bf{V}}_{eq}^{0}}{{\bf{{\tilde{n}}}}_{eq}}$ and ${{\bf{{\tilde{n}}}}_{eq}} \sim \mathcal{CN} \left( {0,\tau {N_0}{{\bf{I}}_{{N_t}}}} \right)$.
\begin{proof}
Please refer to Appendix \ref{proof:theo:channel_estimation}.
\end{proof}
\end{theorem}

{\emph{Remark 1:}} The basic intuition behind Theorem \ref{theo:channel_estimation} is that when $T \rightarrow \infty $ and $N_t \rightarrow \infty$,
each channel tends to be an eigenvector of the received signal matrix. As a result, we project the received signal matrix along
the eigenspace which corresponds to the desired users' channel. In this case, the impact of the strong active
attack can be effectively eliminated, cf. (4).

{\emph{Remark 2:}} It should be noted that when $N_t \rightarrow \infty$, ${P_e} {\rm{tr}}\left( {{\bf{R}}_{E,T}^0} \right) > {P_0} {\rm{tr}} \left({\mathbf {R}_{0k}^0} \right)
 > P_l {\rm{tr}} \left({\mathbf {R}_{lk}^l} \right)$ is enough to distinguish the eavesdropper, the users in the cell of interest, and the users in
 the other cells as shown in Appendix A. However, when $N_t$ is large but not infinite, we require
  ${P_e} {\rm{tr}}\left( {{\bf{R}}_{E,T}^0} \right) \gg {P_0} {\rm{tr}} \left({\mathbf {R}_{0k}^0} \right) \gg P_l {\rm{tr}} \left({\mathbf {R}_{lk}^l} \right)$ to achieve
  a good secrecy performance under the pilot contamination attack.

{\bl {\emph{Remark 3:}} In Theorem \ref{theo:channel_estimation}, we assume that
the coherence time of the channel is significantly larger than
the symbol duration \cite{Muller2014JSTSP}. This assumption can be justified based on the expression
for the coherence time in \cite[Eq. (1)]{Muller2014JSTSP}. For typical speeds of mobile users and typical symbol durations,
the coherence time spans several
hundred symbol durations.

{\emph{Remark 4:}} The simulation results in Section V indicate that a sufficient
power gap between ${P_0}$ and $P_e$ can  guarantee a good secrecy performance
when the number of transmit antennas and the coherence time of the channel
are large but finite. We note that allocating a high power to
 the desired users to combat a strong active attack is not necessary. In contrast,
a large gap  between ${P_0}  {\rm{tr}} \left({\mathbf {R}_{0k}^0} \right) $ and ${P_e} {\rm{tr}}\left( {{\bf{R}}_{E,T}^0} \right) $
is essential to approach the channel estimation result in Theorem \ref{theo:channel_estimation}.
This implies that  \emph{decreasing} the transmit power of the desired users  can be an effective
strategy to ensure secure transmission  under a strong active attack.
As shown in Figure 5 and Figure 6 in Section V, as long as $\rho = P_e/(P_0K)$ is larger than $0$ dB, the proposed design
is able to achieve a good secrecy performance in all considered scenarios.}

Based on Theorem \ref{theo:channel_estimation}, in the next section,  we can design the precoders
for downlink transmission.

\section{Downlink Transmission}
In this section, we consider the downlink transmission phase.
We assume that the BSs in all $L + 1$ cells perform  channel
estimation according to Theorem 1 by replacing ${\widehat {\bf{h}}_{eq,0k}}$,
${{\bf{h}}_{eq,0k}}$, $P_0$, and ${{\bf{V}}_{eq}^{0}}$
by ${\widehat {\bf{h}}_{eq,lk}}$, ${{\bf{h}}_{eq,lk}}$, $P_l$, and
${{\bf{V}}_{eq}^{l}}$, respectively. Then, the $l$th BS designs the transmit
signal as follows
\begin{align}\label{eq:xl}
{{\bf{x}}_l} = \sqrt P \sum\limits_{k = 1}^K {{{\bf{t}}_{lk}}{s_{lk}}}, \quad  l = 0, \cdots, L,
\end{align}
where $P$ is the downlink transmission power,
${{\bf{t}}_{lk}} = \left(\mathbf{V}_{eq}^{l}\right)^H\frac{{{{{\bf{\hat h}}}_{eq,lk}}}}{{\left\| {{{{\bf{\hat h}}}_{eq,lk}}} \right\|}}$,
and $s_{lk}$ is the downlink transmitted signal for the $k$th user in the $l$th cell.

{\bl We note that unlike for the scheme in \cite{Wu2016TIT}, for the proposed precoder
design, the base station does not need to know the full statistical channel state information of the eavesdropper.
As long as ${P_e} {\rm{tr}}\left( {{\bf{R}}_{E,T}^0} \right) \gg {P_0} {\rm{tr}} \left({\mathbf {R}_{0k}^0} \right) \gg P_l {\rm{tr}} \left({\mathbf {R}_{lk}^l} \right)$ holds,
the base station can identify the relevant columns in $\frac{1}{{T{N_t}}}{\mathbf{Y}_0}{{\mathbf{Y}_0^H}} = \left[ {{{\bf{v}}_1},
\cdots ,{{\bf{v}}_{{N_t}}}} \right]{\bf{\Sigma }}{\left[ {{{\bf{v}}_1}, \cdots ,{{\bf{v}}_{{N_t}}}} \right]}$ for computation
of $\mathbf{V}_{eq}^{0}$. For the legitimate users in the cell of interest, the BS needs to
know their covariance matrices ${\bf{R}}_{0k}^0$ to perform  channel estimation, see Theorem 1.}

Because each user in the cell of interest has the risk of being eavesdropped, based on \cite{Geraci2012TCom} and \cite{Wu2016TIT},
the achievable ergodic (the codewords are sent over a large number of fading blocks)
secrecy sum-rate
can be expressed as
\begin{align}\label{eq:sum_rate}
R_{\rm sec} = \sum\limits_{k = 1}^K  \left[R_k - C_k^{\rm eve} \right]^{+}
\end{align}
where $R_k$ and  $C_k^{\rm eve}$ denote an achievable ergodic rate between the BS and the $k$th user and the
ergodic capacity between the BS and the eavesdropper seeking to decode the information of the
$k$th user, respectively.

The received signal ${y_{0k}}$ at the $k$th user in the cell of interest is given by
\begin{align}\label{eq:y0k}
{y_{0k}} & = \sum\limits_{l = 0}^L {{{\left( {{\bf{h}}_{lk}^0} \right)}^H}{{\bf{x}}_l}}  + {{n}_d} \nonumber \\
 & = \sqrt P {\left( {{\bf{h}}_{0k}^0} \right)^H} \left(\mathbf{V}_{eq}^{0}\right)^H \frac{{{{{\bf{\hat h}}}_{eq,0k}}}}{{\left\| {{{{\bf{\hat h}}}_{eq,0k}}} \right\|}}{s_{0k}} \nonumber \\
& + \sqrt P {\left( {{\bf{h}}_{0k}^0} \right)^H}\left(\mathbf{V}_{eq}^{0}\right)^H \sum\limits_{t = 1,t \ne k}^K {\frac{{{{{\bf{\hat h}}}_{eq,0t}}}}{{\left\| {{{{\bf{\hat h}}}_{eq,0t}}} \right\|}}{s_{0t}}}  \nonumber \\
&  + \sqrt P \sum\limits_{l = 1}^L {{{\left( {{\bf{h}}_{lk}^0} \right)}^H}\left(\mathbf{V}_{eq}^{l}\right)^H\sum\limits_{t = 1}^K {\frac{{{{{\bf{\hat h}}}_{eq,lt}}}}{{\left\| {{{{\bf{\hat h}}}_{eq,lt}}} \right\|}}{s_{lt}}} }  + {{n}}_d,
\end{align}
where ${{n}}_d \sim \mathcal{CN} \left( {{0}, N_{0d}} \right)$ is the noise affecting
the received downlink signal.

The achievable ergodic rate $R_k$ is given by
\begin{align}\label{eq:Rk_lower}
{{R}_k} =   E\left[ \log \left( {1 + {\gamma_k}} \right)\right],
\end{align}
where
\begin{align}\label{eq:gamma_k}
& {\gamma _k} = \frac{{{{\left| { {g_{0k,k}^0}} \right|}^2}}}{{{N_{0d}}  + \sum\limits_{t = 1,t \ne k}^K { {{{\left| {g_{0t,k}^0} \right|}^2}} + \sum\limits_{l = 1}^L {\sum\limits_{t = 1}^K { {{{\left| {g_{lt,k}^0} \right|}^2}}} } } }},
\end{align}
and $g_{lt,k}^0 = \sqrt P {\left( {{\bf{h}}_{lk}^0} \right)^H}\left(\mathbf{V}_{eq}^{l}\right)^H\frac{{{{{\bf{\hat h}}}_{eq,lt}}}}{{\left\| {{{{\bf{\hat h}}}_{eq,lt}}} \right\|}}$.

The ergodic capacity of the eavesdropper for decoding the information intended
for user $k$, $C_k^{\rm eve}$, is given by\footnote{It should be noted that here we consider the practical scenario where the eavesdropper
is not able to decode and cancel the signals of the intra-cell
and inter-cell users from the received signal. For a more pessimistic setting,
where the eavesdropper has access
to the data of all intra-cell and inter-cell interfering users, we can also obtain
a lower bound on the ergodic secrecy rate as in \cite{Wu2016TIT}. However, the two expressions exhibit no difference as far as
 the subsequent analysis is concerned since the eavesdropper's rate
will be suppressed to zero based on the proposed data-aided transmission scheme.}
\begin{align}\label{eq:C_eve_k}
C_k^{{\rm{eve}}} =  E\left[{{{\log }_2}\left( {1 + P{{\left( {{{\bf{t}}_{0k}}} \right)}^H}{\bf{H}}_e^0{{\bf{Q}}_k^{ - 1}}{{\left( {{\bf{H}}_e^0} \right)}^H}{{\bf{t}}_{0k}}} \right)}\right],
\end{align}
where
\begin{align}\label{eq:g_eve_k}
{{\bf{Q}}_k} & = {\left( {{\bf{H}}_e^0} \right)^H}\sum\limits_{p = 1,p \ne k}^K {{{\bf{t}}_{0t}}{{\left( {{{\bf{t}}_{0k}}} \right)}^H}{\bf{H}}_e^0} \nonumber \\
 & + \sum\limits_{l = 1}^L {{{\left( {{\bf{H}}_e^l} \right)}^H}\sum\limits_{k = 1}^K {{{\bf{t}}_{lk}}{{\left( {{{\bf{t}}_{lk}}} \right)}^H}} } {\bf{H}}_e^l + {N_{{0d}}}{{\bf{I}}_{{N_e}}}.
\end{align}

Based on (\ref{eq:sum_rate}), (\ref{eq:Rk_lower}), and (\ref{eq:C_eve_k}),
we obtain the following theorem.

\begin{theorem}\label{theo:achievable_rate}
For the considered multi-cell multi-user massive MIMO system,
an asymptotic achievable secrecy sum-rate for the transmit signal design in (\ref{eq:xl})
is given by
\begin{align}\label{eq:sum_rate_lower}
R_{\rm sec,\, ach} \mathop  \to \limits^{{N_t} \to \infty } \sum\limits_{k = 1}^K \log \left( {1 + {{\bar{\gamma}}_k}} \right),
\end{align}
where
\begin{align}\label{eq:gamma_k_bar}
{{\bar{\gamma}}_k} & = \frac{{Pa_{0k,2}^0}}{{a_{0k,1}^0\left( {{N_{0d}}  + P\sum\limits_{t = 1,t \ne k}^K {\frac{{b_{0t,k}^0}}{{a_{0t,1}^0}} + P\sum\limits_{l = 1}^L {\sum\limits_{t = 1}^K {\frac{{c_{lt,k}^l}}{{a_{lt,1}^l}}} } } } \right)}},\\
a{_{lt,1}^l} & = {P_0}\tau \left( {{P_0}\tau \left[ {\sum\limits_{p = 1,p \ne t}^K {{{\left[ {{\bf{A}}_{{{l}t},{{4}}}^l} \right]}_{pp}}{\rm tr}\left( {{\bf{R}}_{lt}^l{\bf{R}}_{lp}^l} \right) }}\right. } \right. \nonumber \\
& \left.{\left.{{  + {{\left[ {{\bf{A}}_{{{l}}t,{{4}}}^l} \right]}_{tt}}{\rm tr}^2 \left( {{\bf{R}}_{lt}^l} \right)} } \right]{{ + }}{N_0}\sum\limits_{p = 1}^K {{{\left[ {{\bf{A}}_{{{l}}t,{{4}}}^l} \right]}_{pp}}{\rm tr}\left( {{\bf{R}}_{lk}^l} \right)} } \right), \label{eq:gamma_k_bar_alt0} \\
{\bf{A}}_{{{l}}t,{{4}}}^l  & = {\bf{A}}_{{{l}}t,{{3}}}^l {\rm diag}\left( {\rm tr} \left({{\bf{R}}_{l1}^l}\right), \cdots, \left({{\bf{R}}_{lK}^l} \right)\right), \\ {\bf{A}}_{{{lt}},{{3}}}^l & = {\left( {{\bf{A}}_{{{lt}},{{1}}}^l} \right)^{{2}}}{\left( {{\bf{A}}_{{{lt}},{{2}}}^l} \right)^{{2}}}, \ {\bf{A}}_{{{lt}},{{2}}}^l = {\left( {{N_0}{{\bf{I}}_K} + \tau {P_0}{\bf{A}}_{{{l t}},{{1}}}^l} \right)^{ - 1}},  \label{eq:gamma_k_bar_A03} \\
 {\bf{A}}_{{{l}}t,{{1}}}^l &  =  \nonumber \\
 & \hspace{-0.8cm} {\rm diag}\left( {\frac{{{\rm{tr}}\left( {{\bf{R}}_{l1}^l{\bf{R}}_{lt}^l} \right)}}{{{\rm{tr}}\left( {{\bf{R}}_{l1}^l} \right)}}, \cdots, \frac{{{\rm{tr}}\left( {{{\left( {{\bf{R}}_{lt}^l} \right)}^2}} \right)}}{{{\rm{tr}}\left( {{\bf{R}}_{lt}^l} \right)}}, \cdots, \frac{{{\rm{tr}}\left( {{\bf{R}}_{lt}^l{\bf{R}}_{lK}^l} \right)}}{{{\rm{tr}}\left( {{\bf{R}}_{lK}^l} \right)}}} \right),  \nonumber \\
 &  l = 0,1,\cdots,L, \quad t =1,\cdots,K,  \label{eq:gamma_k_bar_A01}
\end{align}
\begin{align}
a{_{{{0}}k,2}^0}  & = {P_0}\left( {{P_0}{\tau ^{{2}}}{{\left( {{{\left[ {{\bf{A}}_{{{0k}},{{5}}}^0} \right]}_{kk}} {{\rm tr}^{{2}}}\left( {{\bf{R}}_{0k}^0} \right) } \right. } }}\right.  \nonumber \\
& \left.{{{\left. {+ \sum\limits_{t = 1,t \ne k}^K {{{\left[ {{\bf{A}}_{{{0 k}},{{5}}}^0} \right]}_{tt}}{\rm tr}\left( {{\bf{R}}_{0k}^0{\bf{R}}_{0t}^0} \right)} } \right)}^{{2}}}} \right) \nonumber \\
& {{ + }}{P_0}{N_0}\tau \left( {{{\left[ {{\bf{A}}_{{{0k}},{{4}}}^0} \right]}_{kk}}{\rm tr}^2\left( {{\bf{R}}_{0k}^0} \right) }\right. \nonumber \\
& + \left. {\sum\limits_{t = 1,t \ne k}^K {{{\left[ {{\bf{A}}_{{{0}}k,{{4}}}^0} \right]}_{tt}}{\rm tr}\left( {{\bf{R}}_{0k}^0{\bf{R}}_{0t}^0} \right)} } \right), \ k =1,\cdots,K, \\
 {\bf{A}}_{{{lt}},{{5}}}^l & = {\bf{A}}_{{{lt}},{{2}}}^l{\bf{A}}_{{{lt}},{{1}}}^l {\rm diag}\left( {\rm tr} \left({{\bf{R}}_{01}^0}\right), \cdots, \left({{\bf{R}}_{0K}^0} \right)\right) \nonumber \\
 & = \mathbf{A}_{{{lt}},{{2}}}^l{\bf{A}}_{{{lt}},{{1}}}^l \mathbf{R}_l, \quad l = 0,1,\cdots,L, \quad t =1,\cdots,K, \\
b_{{{0}}t{{,}}k}^0 &  = {\left( {{P_0}\tau {\rm{tr}}\left( {{\bf{R}}_{0k}^0} \right){{\left[ {{\bf{A}}_{{{0t}},{{5}}}^0} \right]}_{kk}}} \right)^{{2}}}{\rm{tr}}\left( {{\bf{R}}_{0k}^0{\bf{R}}_{0t}^0} \right) \nonumber \\
& + {P_0}\tau {N_0}{{}}{{{\rm tr}}^{{3}}}\left( {{\bf{R}}_{0k}^0} \right){\left( {{{\left[ {{\bf{A}}_{{{0t}},{{5}}}^0} \right]}_{kk}}} \right)^{{2}}}, \\
 c_{lt,{{k}}}^l & {{ = }}{P_I}\tau {N_0}\left( {\sum\limits_{p = 1}^K {\left[ {{\bf{A}}_{lt,{{5}}}^l} \right]_{pp}^2 {\rm tr}\left( {{\bf{R}}_{lk}^0{\bf{R}}_{lp}^l} \right){\rm tr}\left( {{\bf{R}}_{lp}^l} \right) }} \right. \nonumber \\
 & \left.{{+ \sum\limits_{p = 1}^K {\sum\limits_{m = 1,m \ne p}^K {{{\left[ {{\bf{A}}_{lt,{{5}}}^l} \right]}_{pp}}{{\left[ {{\bf{A}}_{lt,{{5}}}^l} \right]}_{mm}} {\rm tr} \left( {{\bf{R}}_{lp}^l{\bf{R}}_{lk}^0{\bf{R}}_{lm}^l} \right)} } } } \right).
\end{align}
\begin{proof}
Please refer to Appendix \ref{proof:theo:achievable_rate}.
\end{proof}
\end{theorem}

{\emph{Remark 5:}} Theorem \ref{theo:achievable_rate} is a unified expression which
is valid for arbitrary $K$ and $L$ and general correlated channels.
Also, Theorem \ref{theo:achievable_rate}
indicates that when $N_t$ tends to infinity,
the impact of the active attack from the eavesdropper vanishes
if the proposed joint uplink and downlink transmission approach is adopted.

{\emph{Remark 6:}} Although the rigorous theoretical analysis in Theorem \ref{theo:achievable_rate} requires
that both $N_t$ and $T$ tend to infinity, our
simulation results in Section \ref{sec:simulation} indicate that
even for short packet communication (e.g., $T =128$) and a finite number of transmit antennas
(e.g., $N_t =64$), the proposed  joint uplink and downlink transmission approach is still able
to provide a good secrecy performance under a strong pilot contamination
attack.

{\emph{Remark 7:}} It is important to note that the proposed
scheme does not require joint channel estimation and data detection or
an additional pilot sequence hopping mechanism.
In fact, the uplink data only has to be exploited
to generate  matrix $\mathbf{V}_{eq}^{0}$.
Then,  simply projecting the transmit signal along the eigenspace,
$\mathbf{V}_{eq}^{0}$, is enough to effectively combat the strong pilot contamination
attack without any further computational operations or extra resources.

\section{The I.I.D. Fading Case}
In order to obtain more insightful results, in this section,
we analyze the
asymptotic achievable secrecy rate of massive MIMO systems for
the i.i.d. fading case. For general multi-cell multi-user
massive MIMO systems, we have the following theorem.

\begin{theorem}\label{theo:achievable_rate_iid}
For  multi-cell multi-user massive MIMO systems with i.i.d. fading where $\mathbf{R}_{lk}^p = \beta _{lk}^p \mathbf{I}_{N_t}$\footnote{It should be noted
that here we do not require the channels of the eavesdropper to be i.i.d., i.e.,
$\mathbf{R}^l_{E,T}$ and $\mathbf{R}^l_{E,R}$ can be arbitrary matrices.},
the asymptotic achievable secrecy sum-rate for the transmit signal design in (\ref{eq:xl})
is given by
\begin{align}\label{eq:sum_rate_lower_iid}
R_{\rm sec,\, iid} \mathop  \to \limits^{{N_t} \to \infty }   \sum\limits_{k = 1}^K \log \left( {1 + {\bar{\gamma}_{k,{\rm iid}}}} \right),
\end{align}
where
\begin{align}\label{eq:gamma_k_bar_iid}
{\bar{\gamma}_{k, {\rm iid}}} & = \frac{P{{a_{1k,{\rm iid}}}}}{{N_{0d}}  +  P \sum\limits_{t = 1,t \neq k}^K {a_{2kt,{\rm iid}}}
 +  P \sum\limits_{l = 1}^L {\sum\limits_{t = 1}^K {a_{3ktl,{\rm iid}}}}},\\
 {a_{1k,{\rm{iid}}}} & = \nonumber \\
 & \hspace{-1cm} \frac{{{P_0}\tau {{\left( {\beta _{0k}^0{N_t} + \beta _{0k}^0\left( {K - 1} \right)} \right)}^2} + {N_0}\left( {{N_t}\beta _{0k}^0 + \left( {K - 1} \right)\beta _{0k}^0} \right)}}{{{P_0}\tau \beta _{0k}^0\left( {{N_t} + K - 1} \right) + K{N_0}}}, \\
 {{{a}}_{2kt,{\rm{iid}}}} & = \frac{{{P_0}\tau {N_t}\beta _{0k}^0\beta _{0t}^0 + {N_0}{N_t}\beta _{0k}^0}}{{\left( {{P_0}\tau \beta _{0t}^0\left( {{N_t} + K - 1} \right) + K{N_0}} \right)}}, \\
 {a_{3ktl,{\rm{iid}}}} & =  \frac{{{P_I}\tau {N_t}\beta _{lk}^0\beta _{lt}^0 + \beta _{lk}^0{N_0}\left( {K{\rm{ + }}\frac{{K\left( {K - 1} \right)}}{{{N_t}}}} \right)}}{{{P_I}\tau \beta _{lt}^l\left( {{N_t} + K - 1} \right) + K{N_0}}}.
\end{align}
\begin{proof}
The theorem can be proved by substituting $\mathbf{R}_{lk}^p = \beta _{lk}^p \mathbf{I}_{N_t}$ into (\ref{eq:sum_rate_lower})
and performing some simplifications.
\end{proof}
\end{theorem}

{\emph{Remark 8:}} Theorem \ref{theo:achievable_rate_iid} indicates that the secrecy
rate is a monotonically increasing function of the signal-to-noise ratio (SNR)  ${\rm SNR} = P/N_{0d}$  even
in the presence of an active eavesdropper. This behaviour is
in  sharp contrast with the scheme proposed in \cite[Theorem 3]{Wu2016TIT},
for which the secrecy rate  decreases
for increasing SNR in the high SNR regime if all the available  power at the BS is allocated to the information-carrying
signals. For the proposed data-aided secure massive MIMO transmission,
$\mathbf{V}_{eq}^{0}$  naturally forms an asymptotic orthogonal space
to the eavesdropper's channel. As a result,
the null space of the transmit correlation
matrix of the eavesdropper's channel \cite{Wu2016TIT} is not essential anymore to combat the strong pilot contamination attack.
The proposed joint uplink and downlink transmission scheme can guarantee  reliable secure
communication even for i.i.d. fading channels.

To provide more insights into  the impact of the proposed scheme on secure communication
in massive MIMO systems,
we simplify the system model further to the single-cell
single-user case. Based on Theorem \ref{theo:achievable_rate_iid} and  \cite[Corollary 1]{Evans2000TIT},
we have the following  corollary.

\begin{corollary}\label{coro:achievable_rate_iid}
For a single-cell single-user system ($L = 0, K = 1$) with i.i.d. fading,
the asymptotic achievable secrecy rate for the transmit signal design in (\ref{eq:xl})
is given by
\begin{align}\label{eq:R_iid_single}
R_{\rm sec,\,iid,\,single} = \log \left( {1 + \frac{{P{N_t}\beta _{01}^0}}{{{N_{0d}}}}} + {o}(N_t) \right).
\end{align}
\begin{proof}
By setting $L=0$, $K=1$ in (\ref{eq:sum_rate_lower_iid}) and considering the asymptotic case, we obtain
(\ref{eq:R_iid_single}).
\end{proof}
\end{corollary}

{\emph{Remark 9:}}  The asymptotic secrecy rate in (\ref{eq:R_iid_single})
grows logarithmically with the number of transmit antennas. This is identical
to the growth rate of the asymptotic rate for single-user massive MIMO systems without eavesdropper.
This is in sharp contrast to the conclusions in \cite[Theorem 9]{Wu2016TIT}, where
secure communication is unachievable for the single-cell single-user i.i.d. fading
case for  high pilot contamination attack powers. Corollary \ref{coro:achievable_rate_iid}
reveals that by exploiting  the uplink transmission data, we can find a promising
solution that facilitates secure communication for i.i.d. fading massive MIMO systems with active eavesdropper.

\section{Numerical Results}\label{sec:simulation}
In this section, we present numerical results to evaluate the proposed scheme and the obtained analytical results.
To the best of our knowledge, although there has been a large amount of research on active eavesdropping in the past few years \cite{Basciftci2017,Wang2018JSAC,Do2018TIFS,Tugnait2018TCom},
most of these schemes require some additional pilot signal protocol which is unknown to the eavesdropper
to combat the pilot contamination attack. These schemes are not compatible with
the assumptions in our paper, where the eavesdropper perfectly knows the desired users'
exact pilot signal structure for each transmission.
Hence, we compare the proposed scheme with
matched filter precoding and AN generation
(we refer to this design as MF-AN scheme) and the null space scheme
in \cite{Wu2016TIT}, which both also do
not require  a modified pilot protocol.

We define  ${\rm SNR} = P/N_{0d}$ and set $P_0 =P_1 = ... = P_L$.
Also, we define\footnote{It should be noted that the eavesdropper attacks the $K$ users simultaneously. For each user, the pilot contamination attack power is $P_e/K$. Hence, we define $\rho = P_e/(P_0K)$.}$\rho = P_e/(P_0K)$.
We consider both correlated fading channels and i.i.d. fading
channels. For correlated fading channels,  a uniform linear array is used at the BS with
half a wavelength antenna spacing. The angle of arrival (AoA) interval is $\mathcal{A} = [-\theta_{b},\theta_{b}]$.
The channel power angle spectrum is modeled as a truncated Laplacian distribution
\cite{Cho2010}
\begin{align} \label{eq:p_theta}
p\left( \theta  \right) = \frac{1}{{\sqrt 2 \sigma \left( {1 - {e^{ - \sqrt 2 \pi /\sigma }}} \right)}}{e^{\frac{{ - \sqrt 2 \left\| {\theta  -  \bar{\theta}  } \right\|}}{\sigma }}},
\end{align}
where $\sigma$ and $\bar\theta$ are the angular spread  and the mean  AoA of the channel, respectively.
The angular spread $\sigma$  in (\ref{eq:p_theta}) is assumed to be identical for the channels of all  users and the eavesdropper and
is set to be $\sigma = \pi/2$. The mean channel AoAs, $\bar\theta$, of all  users and the eavesdropper in (\ref{eq:p_theta}) are generated at random.
Based on \cite[Eq. (3.14)]{Cho2010}, we generate
the channel transmit correlation matrices of all users, $\mathbf{R}^p_{lk}$, and the eavesdropper, $\mathbf{R}^l_{E,T}$.
Moreover, we impose a channel power normalization such that the trace of the channel transmit correlation matrix
 between a user
and the BS in its own cell and between a user
and the BSs in the other cells are equal to $N_t$ and $\beta N_t$, respectively.
We set $\beta = 0.1$.  The receive correlation matrices of the eavesdropper, $\mathbf{R}^l_{E,R}$, are generated using the exponential correlation model
 $ \left[\mathbf{R}^l_{E,R} \right]_{i,j} = \varphi^{| i - j |},  \varphi \in (0,1)$, where $\varphi$ is generated at random.
For i.i.d. fading channels, we set $\mathbf{R}^l_{lk} = \beta_{lk}^l \mathbf{I}_{N_t}$, $\mathbf{R}^p_{lk} = \beta_{lk}^p \mathbf{I}_{N_t}$,
$\mathbf{R}^l_{E,T} = \beta_{e}^l \mathbf{I}_{N_t}$, and $\mathbf{R}^l_{E,R} = \mathbf{I}_{N_e}$.
Here, we set  $\beta_{lk}^l = 1$ and $\beta_{lk}^p = 0.1$, $k = 1,\cdots,K$, $l = 1,\cdots,L$, $p = 1,\cdots,L, p \neq l$.
Also, we set $\beta_{e}^0 = 1$ and $\beta_{e}^l = 0.1$, $l = 2,\cdots,L$. Throughout this section, we assume $N_e = 2$.
Table \ref{tab:simulations_2} summarizes the values of main parameters used in the simulations in each figure.

\begin{table*}[!t]
\captionstyle{center}
\centering
{\bl
\caption{Values of main parameters used in simulations.}
\label{tab:simulations_2}
\begin{tabular}{|c|c|c|c|c|c|c|}
\hline
 Parameter   &         Fig. 2   &   Fig. 3   &           Fig. 4 &           Fig. 5&           Fig. 6  &           Fig. 7 \\ \hline
   $L$    &  3   &   3   &  0 &  3 &  3 &  3  \\
\hline
   $K$   & 5    &  5    & 1  & 5  &  5 &  5   \\
\hline
   $N_t$   &   128   & 128, 64     & 64--644 & 128  & 128   & 128   \\
\hline
   $T$   & 128--1024    &  1024     & 16 $N_t$  & 128, 1024  & 128, 1024  &  1024  \\
\hline
    $\tau$   &  64   &   64   &  64  &  8, 64 &   8, 64 & 8   \\
\hline
   $P_0$   &  1, 10, 100   &  10     & 10 & 10  & 10 & 10  \\
\hline
   $N_0$   &    1 &   1   & 1 & 1  & 1  & 1   \\
\hline
  $\rho$   &   1 dB  &   20 dB   & 20 dB   &  0--20 dB   & 0--20 dB    & 10 dB  \\
\hline
${\rm SNR}$   &  None   & -20--26 dB     & 10, 20 dB &  20 dB & 20 dB &  15 dB  \\
\hline
 $\theta_{b}$   &   $\pi$  &  $\pi$    &  None   &  $\pi$  &  None &  $\pi/20$--$\pi$    \\
\hline
$\bar\theta$ &  Random   &   Random  &   None   & Random &  None  &   Random  \\
\hline
 $\sigma$    & $\pi/2$    & $\pi/2$      & None  & $\pi/2$   &  None & $\pi/2$    \\
\hline
$\beta$   & 0.1    &   0.1   &  None &  0.1 & 0.1 & 0.1  \\
\hline
$\beta_{lk}^l $   &  1   &    1  & 1 &  1  & 1  & 1     \\
\hline
$\beta_{lk}^p $   &   0.1   &    0.1  & None &   0.1 &  0.1 &  0.1    \\
\hline
$N_e $   &   2  &   2   & 2 &   2 & 2 &  2  \\
\hline
\end{tabular}}
\vspace*{10pt}

\hrulefill
\end{table*}

{\bl
Figure \ref{MSE} shows the normalized mean squared error (MSE) $ \sum\nolimits_{k = 1}^K \parallel {{{\bf{\hat h}}}_{eq,0k}} -
\mathbf{V}_{eq}^0 \mathbf{h}_{0k}^0 \parallel^2 / \parallel \mathbf{V}_{eq}^0 \mathbf{h}_{0k}^0 \parallel^2 $
versus (vs.) $T$ for the MMSE estimation scheme in Theorem 1 for $L =3$, $K =5$, $N_t = 128$, $\rho = 0$ dB,
$\tau =64$, $N_0 =1$,
correlated fading with $\theta_{b} = \pi$, and different $P_0$.
We observe from Figure \ref{MSE} that the MSE is below $10^{-3}$ for all considered $P_0$  and $T$.
This demonstrates that the proposed data-aided estimation method is an effective approach
to distinguish the  actual channel of each user from the eavesdropper's channel under the pilot contamination attack.

\begin{figure}[!t]
\centering
\includegraphics[width=0.5\textwidth]{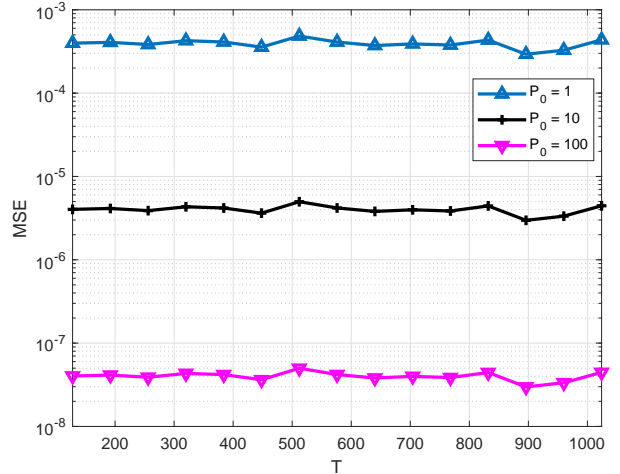}
\caption {\space\space  MSE vs. $T$ for $L =3$, $K =5$, $N_t = 128$, $\rho = 1$ dB, $\tau =64$, $N_0 =1$, correlated fading with $\theta_{b} = \pi$, and different $P_0$.}
\label{MSE}
\end{figure}

}

\begin{figure}[!t]
\centering
\includegraphics[width=0.5\textwidth]{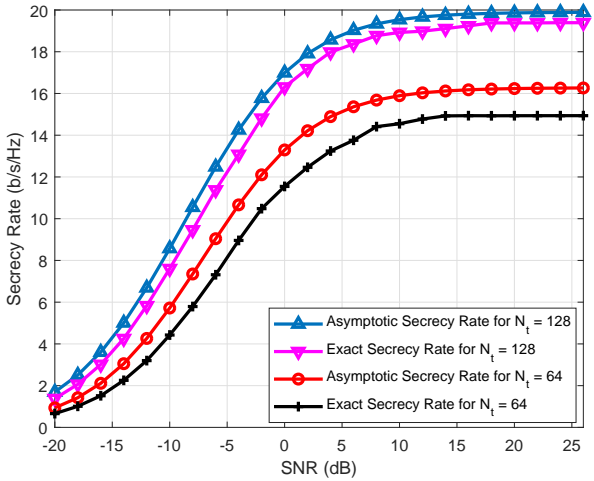}
\caption {\space\space  Secrecy rate vs. SNR for $L =3$, $K =5$,  $T =1024$, $\tau =64$, $\rho$ = 20 dB,  $P_0 =10$, $N_0 =1$, correlated fading with $\theta_{b} = \pi$, and different $N_t$.}
\label{Theory_Mont}
\end{figure}

Figure \ref{Theory_Mont} shows the secrecy rate performance vs. SNR for the proposed scheme
for $L =3$, $K =5$, $T =1024$, $\tau =64$, $\rho$ = 20 dB, $P_0 =10$, $N_0 =1$,
correlated fading with $\theta_{b} = \pi$, and different $N_t$.
 We observe from Figure \ref{Theory_Mont} that the asymptotic secrecy rates
in Theorem \ref{theo:achievable_rate} provide a good approximation
for the exact secrecy rates. The accuracy of the approximation increases with
the number of transmit antennas as expected.  Also, we
observe from Figure \ref{Theory_Mont} that the secrecy rate is a monotonically
increasing function of the SNR even under a strong pilot contamination attack.

\begin{figure}[!t]
\centering
\includegraphics[width=0.5\textwidth]{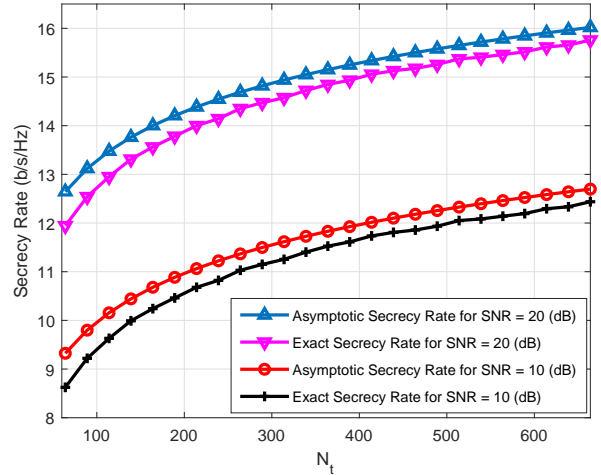}
\caption {\space\space  Secrecy rate vs. $N_t$ for $L =0$, $K =1$, $\rho$ = 20 dB, $P_0 =10$, $N_0 =1$, i.i.d. fading, and different SNRs.}
\label{Theory_Mont_iid}
\end{figure}

Figure \ref{Theory_Mont_iid} shows the secrecy rate performance of the proposed scheme vs. $N_t$
for $L =0$, $K =1$, $\rho$ = 20 dB, $P_0 =10$, $N_0 =1$,
i.i.d. fading, and different SNRs.  We set $\tau = 64$ and $T = 16 N_t$.
We observe from Figure \ref{Theory_Mont_iid} that the secrecy rates
scale logarithmically with the number of transmit antennas,
as predicted by Corollary \ref{coro:achievable_rate_iid}.
Figure \ref{Theory_Mont_iid} also shows that the theoretical secrecy rates
provide good approximations for the exact secrecy rates for the i.i.d. fading case.

\begin{figure}[!t]
\centering
\includegraphics[width=0.5\textwidth]{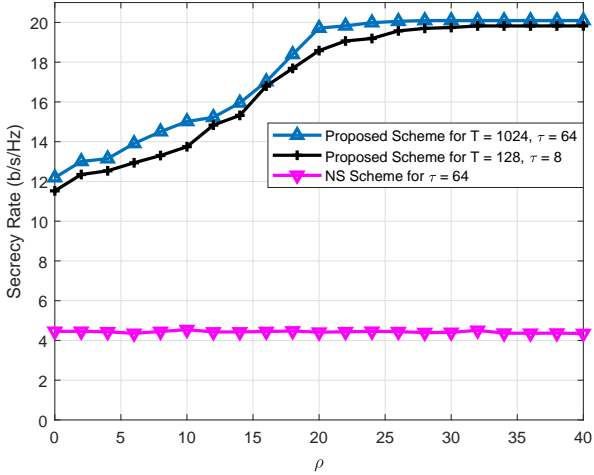}
\caption {\space\space  Secrecy rate vs. $\rho$ for $L =3$, $K =5$, $N_t = 128$, $P_0 =10$, $N_0 =1$, ${\rm SNR} = 20$ dB,
correlated fading with $\theta_{b} = \pi$,  different schemes, and different $T$.}
\label{Fig_Design_correlation}
\end{figure}

Figure \ref{Fig_Design_correlation} shows the exact secrecy rate performance vs. $\rho$ for
the proposed scheme and the null space (NS) scheme \cite{Wu2016TIT}
for $L =3$, $K =5$, $N_t = 128$, $\rho = 20$ dB, $P_0 =10$, $N_0 =1$, ${\rm SNR} = 20$ dB,
correlated fading with $\theta_{b} = \pi$, and different $T$.
We observe from Figure \ref{Fig_Design_correlation} that even for short packet
communication when $T = 128$,  the proposed scheme can achieve
an obvious secrecy performance gain compared to the null space scheme.
Also, we observe from Figure \ref{Fig_Design_correlation} that when
the power of the active attack is strong,  the null space
scheme maintains an almost constant secrecy rate.
However, as $\rho$ increases, the  gap between $P_0 \mathbf{R}_{0k}^0$ and $P_e \Lambda_i$ increases.
Therefore, the secrecy rates of the proposed scheme increase with $\rho$.
Moreover,  Figure \ref{Fig_Design_correlation} reveals that increasing $T$ is beneficial for the
secrecy performance of the proposed scheme.

\begin{figure}[!t]
\centering
\includegraphics[width=0.5\textwidth]{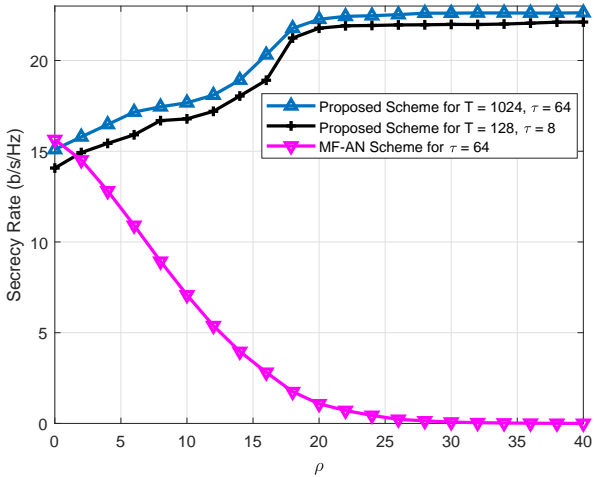}
\caption {\space\space  Secrecy rate vs. $\rho$ for $L =3$, $K =5$, $N_t = 128$, $P_0 =10$, $N_0 =1$, ${\rm SNR} = 20$ dB,
i.i.d. fading,  different schemes, and different $T$.}
\label{Fig_Design_iid}
\end{figure}

Figure \ref{Fig_Design_iid} shows the exact secrecy rate performance vs. $\rho$ for
the proposed scheme and the MF-AN scheme \cite{Wu2016TIT}
for $L =3$, $K =5$, $N_t = 128$, $\rho = 20$ dB, $P_0 =10$, $N_0 =1$, ${\rm SNR} = 20$ dB,
i.i.d. fading, and different $T$. We observe from Figure \ref{Fig_Design_iid} that when
the power of the active attack is strong,  the MF-AN scheme
cannot provide a non-zero secrecy rate.
However, our proposed scheme
performs well in the entire considered range of $\rho$.
When the power of the active attack is close to the power of
the desired user's pilot signal where $\rho$ is small, the asymptotic estimation
error in Theorem 1 increases. As a result,
our proposed scheme  suffers
a slight performance loss comparing to the MF-AN scheme.

\begin{figure}[!t]
\centering
\includegraphics[width=0.5\textwidth]{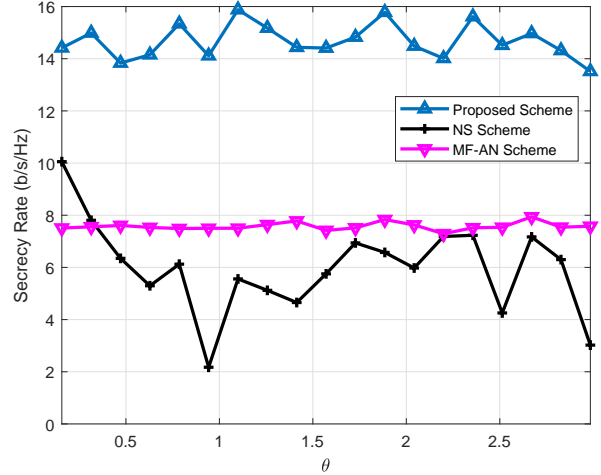}
\caption {\space\space  Secrecy rate vs. $\theta_b$ for $L =3$, $K =5$, $N_t = 128$, $T =1024$, $\tau =64$, $\rho = 10$ dB, $P_0 =10$, $N_0 =1$, ${\rm SNR} = 15$ dB,
correlated fading,  and different schemes.}
\label{Fig_Design_theta}
\end{figure}

Figure \ref{Fig_Design_theta} shows the exact secrecy rate performance vs. $\theta_b$ for
the proposed scheme, the null space scheme, and the MF-AN scheme
for $L =3$, $K =5$, $N_t = 128$, $T =1024$, $\tau =64$, $\rho = 10$ dB, $P_0 =10$, $N_0 =1$, ${\rm SNR} = 15$ dB, and
correlated fading. It should be noted that the rank
of the correlation matrix generated by (\ref{eq:p_theta}) decreases as $\theta_b$ decreases.
This means when $\theta_b$ is small, the channel is highly correlated.
We observe from Figure \ref{Fig_Design_theta} that the proposed
scheme provides the best secrecy performance throughout the entire considered range of $\theta_b$.
Also, we observe from Figure \ref{Fig_Design_theta} that a small value of $\theta_b$ is
beneficial for the null space scheme since the rank of the eavesdropper's transmit correlation matrix
is low.

\section{Conclusions}
In this paper, we have proposed a data-aided
secure transmission scheme for general correlated fading channels and multi-cell
 multi-user massive
MIMO systems  which are under a strong active attack.
We exploit  the received  uplink  data signal
for  joint uplink channel estimation
and secure downlink data transmission.
We show analytically that when the number of transmit
antennas and the length of the data vector both approach infinity,
decreasing (instead of increasing)
the desired user's pilot signal power to
generate an obvious gap to the eavesdropper's pilot
signal power could be an effective approach
for distinguishing the desired user's and the eavesdropper's channels.
Based on this, we propose an effective approach
to eliminate the impact of an active pilot contamination attack.
For the proposed data-aided
secure downlink transmission scheme, we obtain a general asymptotic achievable secrecy
sum-rate expression.
Interestingly, we reveal that for the special case of the classical
MIMO wiretap channel, the obtained expression exhibits
the same scaling
law as the achievable rate of  a single-user massive MIMO system without eavesdropping.
Numerical results validate our theoretical analysis and
demonstrate the effectiveness of the proposed scheme to
mitigate strong active attacks compared  to
 matched filter precoding with artificial noise
generation  and a  null space based scheme.

\appendices

\section{Proof of Theorem \ref{theo:channel_estimation}}\label{proof:theo:channel_estimation}
We define $\boldsymbol{\Omega}_0 = \left[{{\boldsymbol{\omega}}_{1}},\cdots, {{\boldsymbol{\omega}}_{K}}\right]^T$,
$\mathbf{D}_{0} =   \sqrt{P_0} \left[{{{\mathbf{d}}}_{01}},\cdots, {{{\mathbf{d}}}_{0K}}\right]^T$, $\boldsymbol{\Omega}_L = \left[\sqrt{P_1} \boldsymbol{\Omega}_0^T,\cdots, \sqrt{P_L} \boldsymbol{\Omega}_0^T \right]^T$,  $\mathbf{D}_{L}  $ \\
$ = \left[ \sqrt{P_1} {{{\mathbf{d}}}_{11}},\cdots, \sqrt{P_1} {{\mathbf{d}}}_{1K}, \cdots,  \sqrt{P_L} {{\mathbf{d}}}_{L1}, \cdots, \sqrt{P_L} {{\mathbf{d}}}_{LK} \right]^T$, ${{\bf{X}}_0} = \left[\sqrt{P_0}  \boldsymbol{\Omega}_0 \quad \mathbf{D}_{0} \right]$,
${{\bf{X}}_I} =  \left[ \boldsymbol{\Omega}_L \quad \mathbf{D}_{L} \right] $, ${{\bf{X}}_e}  = \left[\sqrt {\frac{{{P_e}}}{{K{N_e}}}} \sum\limits_{k = 1}^K {{{\bf{W}}_k}} \quad \sqrt {\frac{{{P_e}}}{{N_e}}}
\mathbf{A} \right]$.

Based on (\ref{eq:Yp}) and (\ref{eq:Yd}), the received signal $\mathbf{Y}_0$ can be re-expressed as
\begin{align}\label{eq:Y0}
{\mathbf{Y}_0}  = {{\bf{H}}_0}{{\bf{X}}_0} + {{\bf{H}}_I}{{\bf{X}}_I} + {{\bf{H}}_e^0}{{\bf{X}}_e} + {\bf{N}}
\end{align}
where ${\bf{N}} = \left[\mathbf{N}_p^0 \quad \mathbf{N}_d^0 \right]$.

For massive MIMO with correlated fading, when ${N_t} \to \infty $,  we have
\begin{align}
\frac{1}{{{N_t}}}{\bf{H}}_0^H{{\bf{H}}_0} & = \frac{1}{{{N_t}}}\left[ {\begin{array}{*{20}{c}}
{{{\left( {{\bf{h}}_{01}^0} \right)}^H}{\bf{h}}_{01}^0}& \cdots &{{{\left( {{\bf{h}}_{0K}^0} \right)}^H}{\bf{h}}_{01}^0}\\
 \vdots & \ddots & \vdots \\
{{{\left( {{\bf{h}}_{01}^0} \right)}^H}{\bf{h}}_{0K}^0}& \cdots &{{{\left( {{\bf{h}}_{0K}^0} \right)}^H}{\bf{h}}_{0K}^0}
\end{array}} \right]. \label{eq:H00}
\end{align}

Based on \cite[Corollary 1]{Evans2000TIT}, we obtain

\begin{align}
\frac{1}{{{N_t}}}{\bf{H}}_0^H{{\bf{H}}_0} & \mathop  \to \limits^{{N_t} \to \infty } \frac{1}{{{N_t}}}\left[ {\begin{array}{*{20}{c}}
{{\rm{tr}}\left( {{\bf{R}}_{01}^0} \right)}& \mathbf{0} & \cdots & \mathbf{0} \\
\mathbf{0} & \ddots & \ddots & \vdots \\
 \vdots & \ddots & \ddots & \mathbf{0} \\
\mathbf{0} & \cdots & \mathbf{0} &{{\rm{tr}}\left( {{\bf{R}}_{0K}^0} \right)}
\end{array}} \right] \nonumber  \\
 & = \frac{1}{{{N_t}}}{{\bf{R}}_0}. \label{eq:H01}
\end{align}
 Similarly, we have
\begin{align}
\frac{1}{{{N_t}}}{\bf{H}}_I^H{{\bf{H}}_I} & \mathop  \to \limits^{{N_t} \to \infty } \frac{1}{{{N_t}}} \left[ {\begin{array}{*{20}{c}}
{{\rm{tr}}\left( {{\bf{R}}_{11}^0} \right)}& \mathbf{0} & \cdots & \mathbf{0}\\
\mathbf{0} & \ddots & \ddots & \vdots \\
 \vdots & \ddots & \ddots & \mathbf{0} \\
\mathbf{0} & \cdots &\mathbf{0} &{{\rm{tr}}\left( {{\bf{R}}_{LK}^0} \right)}
\end{array}} \right]  \nonumber \\
& = \frac{1}{{{N_t}}}{{\bf{R}}_I}. \label{eq:HI}
\end{align}
Also, based on \cite[Eq. (102)]{Wen2013TIT}, we have
\begin{align}
& \frac{1}{{{N_t}}}{\left[ {{\bf{H}}_e^H {{\bf{H}}_e} } \right]_{{{ij}}}} \nonumber \\
& = \frac{1}{{{N_t}}}{\left( {{\bf{h}}_{E,i}^0} \right)^H}{\bf{h}}_{E,j}^0 \nonumber \\
& = {\bf{e}}_i^H{\left( {{\bf{R}}_{E,R}^0} \right)^{1/2}}{\left( {{\bf{G}}_E^0} \right)^H}{\left( {{\bf{R}}_{E,T}^0} \right)^{1/2}} \nonumber \\
& \hspace{3cm} \times {\left( {{\bf{R}}_{E,T}^0} \right)^{1/2}}{\bf{G}}_E^0{\left( {{\bf{R}}_{E,R}^0} \right)^{1/2}}{\bf{e}}_j^H \nonumber \\
& ={\rm{ tr}} \left( {{\bf{e}}_i^H{{\left( {{\bf{R}}_{E,R}^0} \right)}^{1/2}}{{\left( {{\bf{G}}_E^0} \right)}^H}{{\left( {{\bf{R}}_{E,T}^0} \right)}^{1/2}}
{{\left( {{\bf{R}}_{E,T}^0} \right)}^{1/2}} }\right. \nonumber \\
& \hspace{3cm} \times \left. {{\bf{G}}_E^0{{\left( {{\bf{R}}_{E,R}^0} \right)}^{1/2}}{{\bf{e}}_j}} \right)  \nonumber \\
& \mathop  \to \limits^{{N_t} \to \infty } \frac{1}{{{N_t}}}{\rm{tr}}\left( {{{\left( {{\bf{R}}_{E,R}^0} \right)}^{1/2}}{{\bf{e}}_j}{\bf{e}}_i^H{{\left( {{\bf{R}}_{E,R}^0} \right)}^{1/2}}} \right){\rm{tr}}\left( {{\bf{R}}_{E,T}^0} \right) \nonumber \\
& =  \frac{1}{{{N_t}}}{\left\{ {{\bf{R}}_{E,R}^0} \right\}_{ij}}{\rm{tr}}\left( {{\bf{R}}_{E,T}^0} \right) \label{eq:HE_0}
\end{align}
where ${{\bf{h}}_{E,i}^0}$ denotes the $i$th column of matrix ${{\bf{H}}_e}$,
${{\bf{G}}_E^0} \sim \mathcal{CN}\left(\mathbf{0}, \mathbf{I}_{N_t} \otimes  \mathbf{I}_{N_e}\right)$,
and ${\bf{e}}_i$ is a $N_t \times 1$ vector with the $i$th element being one
and the other elements being zero.

As a result, we have
\begin{align}
\frac{1}{{{N_t}}}{\bf{H}}_e^H{{\bf{H}}_e}\mathop  \to \limits^{{N_t} \to \infty } \frac{1}{{{N_t}}}{\bf{R}}_{E,R}^0{\rm{tr}}\left( {{\bf{R}}_{E,T}^0} \right){\rm{ = }}\frac{1}{{{N_t}}}{{\bf{R}}_E}.  \label{eq:HE_01}
\end{align}

When $ T\rightarrow \infty $, based on \cite[Corollary 1]{Evans2000TIT}, we have
\begin{align}
\frac{1}{T}{{\bf{X}}_0}{\bf{X}}_0^H & \mathop  \to \limits^{T \to \infty } {P_0}{{\bf{I}}_K}, \\
 \frac{1}{T}{{\bf{X}}_I}{\bf{X}}_I^H &\mathop  \to \limits^{T \to \infty } {P_I}{{\bf{I}}_{LK}}, \\
 \frac{1}{T}{{\bf{X}}_e}{\bf{X}}_e^H & \mathop  \to \limits^{T \to \infty } {P_e}{{\bf{I}}_{{N_e}}}.
\end{align}

Then, we obtain (\ref{YY_decom_1}) given at the top of the next page.

 \begin{figure*}[!ht]
\begin{align}
& \frac{1}{{{N_t}T}}{\bf{Y}}{{\bf{Y}}^H}  \mathop  \to \limits^{{N_t} \to \infty ,T \to \infty } \frac{1}{{{N_t}T}}{{\bf{H}}_0}{{\bf{X}}_0}{\bf{X}}_0^H{\bf{H}}_0^H
 + \frac{1}{{{N_t}T}}{{\bf{H}}_I}{{\bf{X}}_I}{\bf{X}}_I^H{\bf{H}}_I^H + \frac{1}{{{N_t}T}}{{\bf{H}}_e}{{\bf{X}}_e}{\bf{X}}_e^H{\bf{H}}_e^H + \frac{{{N_0}}}{{{N_t}}}{{\bf{I}}_{{N_t}}} \nonumber \\
& = \frac{1}{{{N_t}}}\left[ {\begin{array}{*{20}{c}}
{{{\bf{U}}_W}}&{{{\bf{H}}_I}{\bf{R}}_I^{ - 1/2}}&{{{\bf{H}}_e}{\bf{R}}_E^{ - 1/2}}&{{{\bf{H}}_0}{\bf{R}}_0^{ - 1/2}}
\end{array}} \right] \nonumber \\
& \hspace{-2cm} \left[ {\begin{array}{*{20}{c}}
{{N_0}{{\bf{I}}_{{N_t} - M}}}&{\bf{0}}& \cdots &{\bf{0}}\\
{\bf{0}}&{\frac{{{\bf{R}}_I^{1/2}{{\bf{X}}_I}{\bf{X}}_I^H{\bf{R}}_I^{1/2}}}{T} + {N_0}{{\bf{I}}_{\left( {L - 1} \right)K}}}& \ddots & \vdots \\
 \vdots & \ddots &{\frac{{{\bf{R}}_E^{1/2}{{\bf{X}}_e}{\bf{X}}_e^H{\bf{R}}_E^{1/2}}}{T} + {N_0}{{\bf{I}}_{{N_e}}}}&{\bf{0}}\\
{\bf{0}}& \cdots &{\bf{0}}&{\frac{{{\bf{R}}_0^{1/2}{{\bf{X}}_0}{\bf{X}}_0^H{\bf{R}}_0^{1/2}}}{T} + {N_0}{{\bf{I}}_K}}
\end{array}} \right] \left[ {\begin{array}{*{20}{c}}
{{\bf{U}}_W^H}\\
{{\bf{R}}_I^{ - 1/2}{\bf{H}}_I^H}\\
{{\bf{R}}_E^{ - 1/2}{\bf{H}}_e^H}\\
{{\bf{R}}_0^{ - 1/2}{\bf{H}}_0^H}
\end{array}} \right]. \label{YY_decom_1}
\end{align}
\vspace*{6pt}

 \hrulefill
\vspace*{4pt}
\end{figure*}

Based on \cite[Corollary 1]{Evans2000TIT}, we obtain (\ref{YY_decom_2}) given at the top of next page
from (\ref{YY_decom_1}).
 \begin{figure*}[!ht]
\begin{align}
 \frac{1}{{{N_t}T}}{\bf{Y}}{{\bf{Y}}^H} & \mathop  \to \limits^{{N_t} \to \infty ,T \to \infty }  \frac{1}{{{N_t}}}\left[ {\begin{array}{*{20}{c}}
{{{\bf{U}}_W}}&{{{\bf{H}}_I}{\bf{R}}_I^{ - 1/2}}&{{{\bf{H}}_e}{\bf{R}}_E^{ - 1/2}}&{{{\bf{H}}_0}{\bf{R}}_0^{ - 1/2}}
\end{array}} \right] \nonumber
\end{align}
\begin{align}
& \left[ {\begin{array}{*{20}{c}}
{{N_0}{{\bf{I}}_{{N_t} - M}}}&{\bf{0}}& \cdots &{\bf{0}}\\
{\bf{0}}&{{P_I}{{\bf{R}}_I} + {N_0}{{\bf{I}}_{LK}}}& \ddots & \vdots \\
 \vdots & \ddots &{{P_{\rm{e}}}{{\bf{R}}_e} + {N_0}{{\bf{I}}_{{N_e}}}}&{\bf{0}}\\
{\bf{0}}& \cdots &{\bf{0}}&{{P_0}{{\bf{R}}_0} + {N_0}{{\bf{I}}_K}}
\end{array}} \right]
\left[ {\begin{array}{*{20}{c}}
{{\bf{U}}_W^H}\\
{{\bf{R}}_I^{ - 1/2}{\bf{H}}_I^H}\\
{{\bf{R}}_E^{ - 1/2}{\bf{H}}_e^H}\\
{{\bf{R}}_0^{ - 1/2}{\bf{H}}_0^H}
\end{array}} \right].  \label{YY_decom_2}
\end{align}
\vspace*{6pt}

 \hrulefill
\vspace*{4pt}
\end{figure*}

Now, we re-write (\ref{YY_decom_2}) as (\ref{YY_decom_3}) given at the top of the next page.
 \begin{figure*}[!ht]
\begin{align}
& \frac{1}{{{N_t}T}}{\bf{Y}}{{\bf{Y}}^H}  \mathop  \to \limits^{{N_t} \to \infty ,T \to \infty } =  \frac{1}{{{N_t}}}\left[ {\begin{array}{*{20}{c}}
{{{\bf{U}}_W}}&{{{\bf{H}}_I}{\bf{R}}_I^{ - 1/2}}&{{{\bf{H}}_e}{\bf{R}}_E^{ - 1/2}{{\bf{U}}_E}}&{{{\bf{H}}_0}{\bf{R}}_0^{ - 1/2}}
\end{array}} \right] \nonumber \\
& \left[ {\begin{array}{*{20}{c}}
{{N_0}{{\bf{I}}_{{N_t} - M}}}&{\bf{0}}& \cdots &{\bf{0}}\\
{\bf{0}}&{{P_I}{{\bf{R}}_I} + {N_0}{{\bf{I}}_{\left( {L - 1} \right)K}}}& \ddots & \vdots \\
 \vdots & \ddots &{{P_{\rm{e}}}{{\boldsymbol{\Lambda }}_{\rm{e}}} + {N_0}{{\bf{I}}_{{N_e}}}}&{\bf{0}}\\
{\bf{0}}& \cdots &{\bf{0}}&{{P_0}{{\bf{R}}_0} + {N_0}{{\bf{I}}_K}}
\end{array}} \right] \left[ {\begin{array}{*{20}{c}}
{{\bf{U}}_W^H}\\
{{\bf{R}}_I^{ - 1/2}{\bf{H}}_I^H}\\
{{\bf{U}}_E^H{\bf{R}}_E^{ - 1/2}{\bf{H}}_e^H}\\
{{\bf{R}}_0^{ - 1/2}{\bf{H}}_0^H}
\end{array}} \right].  \label{YY_decom_3}
\end{align}
\vspace*{6pt}

 \hrulefill
\vspace*{4pt}
\end{figure*}

We define
\begin{align}
\mathbf{U}_Y & = \left[ {\begin{array}{*{20}{c}}
{{{\bf{U}}_W}}&{{{\bf{H}}_I}{\bf{R}}_I^{ - 1/2}}&{{{\bf{H}}_e}{\bf{R}}_E^{ - 1/2}}&{{{\bf{H}}_0}{\bf{R}}_0^{ - 1/2}}
\end{array}} \right]  \label{YYH-B0}
\end{align}
where ${{{\bf{U}}_W}} \in \mathbb{C} {^{N_t \times \left(N_t - M \right)}}$ has orthogonal columns.

Based on (\ref{eq:H01})--(\ref{eq:HE_01}), we know
\begin{align}\label{eq:UY}
\frac{1}{{{N_t}}}  \mathbf{U}_Y^H \mathbf{U}_Y \mathop \to \limits^{{N_t} \to \infty} \mathbf{I}_{N_t}.
\end{align}

From (\ref{YY_decom_1})--(\ref{eq:UY}), we know that for $T \to \infty$, $N_t \to \infty$,
$\mathbf{U}_Y$ is the right singular matrix
of ${\bf{Y}}_0$. Therefore, we obtain
\begin{align}\label{eq:Z}
{\bf{Z}}_{0p} & = \frac{1}{{\sqrt {T{N_t}} }}{\left(\mathbf{V}_{eq}^{0}\right)^H}{\bf{Y}}_p^0 \mathop \to \limits^{{N_t} \to \infty }   \frac{1}{{\sqrt {T{N_t}} }}{\left(\mathbf{V}_{eq}^{0}\right)^H}  \sqrt{P_0}  \boldsymbol{\Omega}_0  {{\bf{X}}_0} \nonumber  \\
& + \frac{1}{{\sqrt {T{N_t}} }}{\left(\mathbf{V}_{eq}^{0}\right)^H}{\bf{N}}_p^0.
\end{align}

Define $\mathbf{z} = {\rm vec} \left( {\mathbf{Z}_{0p}} \right) $, where ${\mathbf{Z}_{0p}}$ is defined
in Theorem \ref{theo:channel_estimation}. From (\ref{eq:Z}), we can re-express the equivalent received signal during the pilot transmission
phase as follows
\begin{align}\label{eq:pilot_phase}
{\bf{z}} = \sqrt {{P_0}} \sum\limits_{t = 1}^K {\left( {{{\boldsymbol{\omega }}_t} \otimes {{\bf{I}}_K}} \right){{\bf{h}}_{eq,0t}}}  + {\bf{n}}
\end{align}
where
\begin{align}\label{eq:noise}
{\bf{n}} = \left[ \begin{array}{l}
{\left( {{\bf{V}}_{eq}^0} \right)^H}{{\bf{n}}_{p1}^{0}}\\
 \vdots \\
{\left( {{\bf{V}}_{eq}^0} \right)^H}{{\bf{n}}_{p\tau}^{0} }
\end{array} \right]
\end{align}
and ${\bf{n}}_{pt}^{0}$ in (\ref{eq:noise}) is the $t$th column of  $\mathbf{N}_p^0$.

Based on (\ref{eq:pilot_phase}),  the MMSE estimate of ${{\bf{h}}_{eq,0k}}$ is given by
\begin{align}
{{{\bf{\hat h}}}_{eq,0k}} & = \sqrt {{P_0}} {{\bf{V}}_{{0}}}{\bf{R}}_{0k}^0{\bf{V}}_{{0}}^H{\left( {{{\boldsymbol{\omega }}_k} \otimes {{\bf{I}}_K}} \right)^H} \nonumber \\
& \hspace{-1cm} \times \left(\! {{N_0}{{\bf{I}}_{K\tau }} + {P_0}\sum\limits_{t = 1}^K  \left( {{{\boldsymbol{\omega }}_k} \otimes {{\bf{I}}_K}} \right)
 {{\bf{V}}_{{0}}}{\bf{R}}_{0t}^0{\bf{V}}_{{0}}^H{{\left( {{{\boldsymbol{\omega }}_k} \otimes {{\bf{I}}_K}} \right)}^H} } \!\right)^{ - 1}{\bf{z}} \nonumber \\
 & = \sqrt {{P_0}} {{\bf{V}}_{{0}}}{\bf{R}}_{0k}^0{\bf{V}}_{{0}}^H{\left( {{N_0}{{\bf{I}}_K} + \tau {P_0}{{\bf{V}}_{{0}}}{\bf{R}}_{0k}^0{\bf{V}}_{{0}}^H} \right)^{ - 1}} \nonumber \\
& \times \left( {\sqrt {{P_0}} \tau {{\bf{h}}_{eq,0k}} + {{\left( {{{\boldsymbol{\omega }}_k} \otimes {{\bf{I}}_K}} \right)}^H}{\bf{n}}} \right). \label{eq:mmse_heq0m_2}
\end{align}
For the noise term in (\ref{eq:mmse_heq0m_2}), we have
\begin{align}
{\left( {{{\boldsymbol{\omega }}_k} \otimes {{\bf{I}}_K}} \right)^H}{\bf{n}} & = \left(\mathbf{V}_{eq}^{0}\right)^H \sum\limits_{t = 1}^\tau  {\omega _{kt}^*}
{{\bf{n}}_{pt}^{0}}
\nonumber  \\
& = \left(\mathbf{V}_{eq}^{0}\right)^H {{\bf{\tilde{n}}}_{eq}} =  \mathbf{n}_{eq} \label{eq:mmse_noise}
\end{align}
where ${\omega _{kt}}$ is the $t$th element of ${\boldsymbol{\omega }}_k$.

Combining (\ref{eq:mmse_heq0m_2}) and (\ref{eq:mmse_noise}) completes the proof.

\section{Proof of Theorem \ref{theo:achievable_rate}}\label{proof:theo:achievable_rate}
From (\ref{YY_decom_2}), we know
\begin{align}
\left({\bf{V}}_{eq}^0\right)^H = {{\bf{H}}_0}{\bf{R}}_0^{ - 1/2} = \left[ {{\bf{h}}_{01}^0, \cdots {\bf{h}}_{0K}^0} \right]{\bf{R}}_0^{ - 1/2}, \label{V0} \\
\left({\bf{V}}_{eq}^l\right)^H = {{\bf{H}}_l} {\bf{R}}_l^{ - 1/2}   = \left[ {{\bf{h}}_{01}^0, \cdots {\bf{h}}_{0K}^0} \right]{\bf{R}}_l^{ - 1/2}.  \label{Vl}
\end{align}

First, we consider
\begin{align}
\left|g_{0k,k}^0\right|^2 =  \frac{{P\left({{\bf{h}}_{0k}^0} \right)^H}\left(\mathbf{V}_{eq}^{l}\right)^H
{{{{{\bf{\hat h}}}_{eq,0k}}}}{{{{{\bf{\hat h}}}_{eq,0k}^H}}}\mathbf{V}_{eq}^{l}{{\bf{h}}_{0k}^0}}{{\left\| {{{{\bf{\hat h}}}_{eq,0k}}} \right\|^2}}. \label{g0kk}
\end{align}

Based on (\ref{eq:h_est}), we have
\begin{align}
{\left\| {{{{\bf{\hat h}}}_{eq,0k}}} \right\|^2} & = {P_0}{\left( {\sqrt {{P_0}} \tau {\bf{V}}_{eq}^0{\bf{h}}_{0k}^0 + {\bf{V}}_{eq}^0{{\tilde {\bf{n}}}_{eq}}} \right)^H} \nonumber \\
& \times {\left(\!{{N_0}{{\bf{I}}_K} + \tau {P_0}{\bf{V}}_{eq}^0{\bf{R}}_{0k}^0{{\left( {{\bf{V}}_{eq}^0} \right)}^H}} \!\right)^{ - 1}}{\bf{V}}_{eq}^0{\bf{R}}_{0k}^0{\left(\! {{\bf{V}}_{eq}^0} \!\right)^H} \nonumber \\
& \times {\bf{V}}_{eq}^0{\bf{R}}_{0k}^0{\left( {{\bf{V}}_{eq}^0} \right)^H}{\left( {{N_0}{{\bf{I}}_K} + \tau {P_0}{\bf{V}}_{eq}^0{\bf{R}}_{0k}^0{{\left(\! {{\bf{V}}_{eq}^0} \!\right)}^H}}\! \!\right)^{ - 1}} \nonumber \\
& \times \left( {\sqrt {{P_0}} \tau  {\bf{V}}_{eq}^0{\bf{h}}_{0k}^0 + {\bf{V}}_{eq}^0{{\bf{{\tilde{n}}}}_{eq}}} \right).
 \label{eq:h_est_norm_2}
\end{align}

Based on (\ref{V0}) and \cite[Corollary 1]{Evans2000TIT}, we have
\begin{align}
& \frac{1}{{{N_t}}} {\bf{V}}_{eq}^0{\bf{R}}_{0k}^0{\left( {{\bf{V}}_{eq}^0} \right)^H}  \\
&  = \frac{1}{{{N_t}}} {\bf{R}}_0^{ - 1/2}{\left[ {{\bf{h}}_{01}^0, \cdots {\bf{h}}_{0K}^0} \right]^H}{\bf{R}}_{0k}^0\left[ {{\bf{h}}_{01}^0, \cdots {\bf{h}}_{0K}^0} \right]{\bf{R}}_0^{ - 1/2}  \nonumber \\
& \mathop  \to \limits^{{N_t} \to \infty } \frac{1}{{{N_t}}}{\bf{R}}_0^{ - 1/2}  {\rm diag}\left( {{\rm{tr}}\left( {{\bf{R}}_{01}^0{\bf{R}}_{0k}^0} \right), \cdots ,{\rm{tr}}\left( {{{\left( {{\bf{R}}_{0k}^0} \right)}^2}} \right) }\right. \nonumber \\
& \hspace{3cm} \left.{, \cdots ,{\rm{tr}}\left( {{\bf{R}}_{01}^0{\bf{R}}_{0K}^0} \right)} \right){\bf{R}}_0^{ - 1/2} \nonumber \\
& = \frac{1}{{{N_t}}} {\rm diag} \! \left(\!{\frac{{{\rm{tr}}\left( {{\bf{R}}_{01}^0{\bf{R}}_{0k}^0} \right)}}{{{\rm{tr}}\left( {{\bf{R}}_{01}^0} \right)}} \cdots \frac{{{\rm{tr}}\left( {{{\left( {{\bf{R}}_{0k}^0} \right)}^2}} \right)}}{{{\rm{tr}}\left( {{\bf{R}}_{0k}^0} \right)}} \cdots \frac{{{\rm{tr}}\left( {{\bf{R}}_{0k}^0{\bf{R}}_{0K}^0} \right)}}{{{\rm{tr}}\left( {{\bf{R}}_{0K}^0} \right)}}}\!\right) \nonumber \\
&  = {\bf{A}}_{{{0}}k,{{1}}}^0.
 \label{eq:h_VRV}
\end{align}

By substituting (\ref{eq:h_VRV}) into (\ref{eq:h_est_norm_2}) and simplifying, we have
\begin{align}
\frac{1}{{{N_t}}}{\left\| {{{{\bf{\hat h}}}_{eq,0k}}} \right\|^2} & = \frac{1}{{{N_t}}}{P_0}{\left( {\sqrt {{P_0}} \tau {\bf{V}}_{eq}^0{\bf{h}}_{0k}^0 + {\bf{V}}_{eq}^0 {{\bf{{\tilde{n}}}}_{eq}} } \right)^H}{\bf{A}}_{{{0k}},{{3}}}^0   \nonumber \\
& \times \left( {\sqrt {{P_0}} \tau {\bf{V}}_{eq}^0 {\bf{h}}_{0k}^0 + {\bf{V}}_{eq}^0 {{\bf{{\tilde{n}}}}_{eq}}} \right) \nonumber \\
& \mathop  \to \limits^{{N_t} \to \infty } \frac{1}{{{N_t}}}{P_0}\left( {{P_0}{\tau ^{{2}}}{{\left( {{\bf{h}}_{0k}^0} \right)}^H} \left( {{\bf{V}}_{eq}^0} \right)^H {\bf{A}}_{{{0k}},{{3}}}^0
{\bf{h}}_{0k}^0 }\right.\nonumber \\
&  \left.{ + {{\left( {{\bf{{\tilde{n}}}}_{eq}} \right)}^H} \left( {{\bf{V}}_{eq}^0} \right)^H {\bf{A}}_{{{0k}},{{3}}}^0  {\bf{V}}_{eq}^0 {{\bf{{\tilde{n}}}}_{eq}}} \right),  \label{eq:eq:h_est_norm_3}
\end{align}
where ${\bf{A}}_{{{0k}},{{3}}}^0$ is defined in (\ref{eq:gamma_k_bar_A03}).

For the first term on the right hand side of (\ref{eq:eq:h_est_norm_3}), we obtain
\begin{align}
& \frac{1}{{{N_t}}}{\left( {{\bf{h}}_{0k}^0} \right)^H} \left( {{\bf{V}}_{eq}^0} \right)^H {\bf{A}}_{{{0k}},{{3}}}^0  {\bf{V}}_{eq}^0    {\bf{h}}_{0k}^0 \nonumber \\
& = \frac{1}{{{N_t}}}{\left( {{\bf{h}}_{0k}^0} \right)^H}\left[ {{\bf{h}}_{01}^0, \cdots {\bf{h}}_{0K}^0} \right]{\bf{R}}_0^{ - 1/2}{\bf{A}}_{{{0k}},{{3}}}^0 \nonumber \\
& \times {\bf{R}}_0^{ - 1/2}{\left[ {{\bf{h}}_{01}^0, \cdots {\bf{h}}_{0K}^0} \right]^H}{\bf{h}}_{0k}^0 \nonumber \\
& {{ = }}\frac{1}{{{N_t}}}\left[ {\sum\limits_{t = 1,t \ne k}^K {{{\left( {{\bf{h}}_{0k}^0} \right)}^H}{\bf{h}}_{0t}^0{{\left[ {{\bf{A}}_{{{0}}k,{{4}}}^0} \right]}_{tt}}{{\left( {{\bf{h}}_{0t}^0} \right)}^H}{\bf{h}}_{0k}^0 }}\right. \\
& \left. {{+ {{\left( {{\bf{h}}_{0k}^0} \right)}^H}{\bf{h}}_{0k}^0{{\left[ {{\bf{A}}_{{{0k}},{{4}}}^0} \right]}_{kk}}{{\left( {{\bf{h}}_{0k}^0} \right)}^H}{\bf{h}}_{0k}^0} } \right],
\end{align}
where ${\bf{A}}_{{{0k}},{{4}}}^0$ is defined in (\ref{eq:gamma_k_bar_A03}).

Based on \cite[Corollary 1]{Evans2000TIT}, we obtain
\begin{align}
& \frac{1}{{{N_t}}}{\left( {{\bf{h}}_{0k}^0} \right)^H} \left( {{\bf{V}}_{eq}^0} \right)^H {\bf{A}}_{{{0k}},{{3}}}^0  {\bf{V}}_{eq}^0    {\bf{h}}_{0k}^0 \nonumber \\
& \mathop  \to \limits^{{N_t} \to \infty } \frac{1}{{{N_t}}}\left[ {\sum\limits_{t = 1,t \ne k}^K {{{\left[ {{\bf{A}}_{{{0k}},{{4}}}^0} \right]}_{tt}}{\rm{tr}}\left( {{\bf{R}}_{0k}^0{\bf{h}}_{0t}^0{{\left( {{\bf{h}}_{0t}^0} \right)}^H}} \right) }}\right. \nonumber \\
& \hspace{4cm} \left. {{+ {{\left[ {{\bf{A}}_{{{0}}k,{{4}}}^0} \right]}_{kk}}{\rm{t}}{{\rm{r}}^2}\left( {{\bf{R}}_{0k}^0} \right)} } \right]  \nonumber \\
& \mathop  \to \limits^{{N_t} \to \infty } \frac{1}{{{N_t}}}\left[ {\sum\limits_{t = 1,t \ne k}^K {{{\left[ {{\bf{A}}_{{{0k}},{{4}}}^0} \right]}_{tt}}{\rm{tr}}\left( {{\bf{R}}_{0k}^0{\bf{R}}_{0t}^0} \right) }} \right. \nonumber \\
& \hspace{4cm} \left. {{+ {{\left[ {{\bf{A}}_{{{0k}},{{4}}}^0} \right]}_{kk}}{\rm{t}}{{\rm{r}}^2}\left( {{\bf{R}}_{0k}^0} \right)} } \right]. \label{eq:norm_first_term}
\end{align}

For the second term on the right hand side of (\ref{eq:eq:h_est_norm_3}), we have
\begin{align}
& \frac{1}{{{N_t}}}{\left( {{\bf{{\tilde{n}}}}_{eq}} \right)^H} \left( {{\bf{V}}_{eq}^0} \right)^H   {\bf{A}}_{{{0k}},{{3}}}^0 {\bf{V}}_{eq}^0  {{\bf{{\tilde{n}}}}_{eq}}  \mathop  \to \limits^{{N_t} \to \infty } \\
& \frac{1}{{{N_t}}}\tau {N_0}{\rm{tr}}\left(\left( {{\bf{V}}_{eq}^0} \right)^H  {\bf{A}}_{{{0}}k,{{3}}}^0  \left( {{\bf{V}}_{eq}^0} \right)  \right) \nonumber \\
&  = \frac{1}{{{N_t}}}\tau {N_0} \nonumber \\
 & \times {\rm{tr}}\left( {\left[ {{\bf{h}}_{01}^0, \cdots {\bf{h}}_{0K}^0} \right]{\bf{R}}_0^{ - 1/2}{\bf{A}}_{{{0}}k,{{3}}}^0{\bf{R}}_0^{ - 1/2}{{\left[ {{\bf{h}}_{01}^0, \cdots {\bf{h}}_{0K}^0} \right]}^H}} \right) \nonumber \\
&  = \frac{1}{{{N_t}}}\tau {N_0}\sum\limits_{t = 1}^K {{{\left[ {{\bf{A}}_{{{0}}k,{{4}}}^0} \right]}_{tt}}{\rm{tr}}\left( {{\bf{h}}_{0t}^0{{\left( {{\bf{h}}_{0t}^0} \right)}^H}} \right)} \nonumber \\
& = \frac{1}{{{N_t}}}\tau {N_0}\sum\limits_{t = 1}^K {{{\left[ {{\bf{A}}_{{{0}}k,{{4}}}^0} \right]}_{tt}}{\rm{tr}}\left( {{\bf{R}}_{0t}^0} \right)}. \label{eq:norm_second_term}
\end{align}

Combining (\ref{eq:eq:h_est_norm_3})--(\ref{eq:norm_second_term}) yields
\begin{align}
\frac{1}{{{N_t}}}{\left\| {{{{\bf{\hat h}}}_{eq,0k}}} \right\|^2}\mathop  \to \limits^{{N_t} \to \infty } \frac{1}{{{N_t}}} a{_{{{0k}},1}^0}, \label{eq:hk_est}
\end{align}
where $a{_{{{0k}},1}^0}$ is defined in (\ref{eq:gamma_k_bar_alt0}).

For the numerator in (\ref{g0kk}), we have (\ref{eq:h0k_v_eq_1}) given at the top of the next page, where ${\bf{A}}_{{{0k}},{{1}}}^0$ and ${\bf{A}}_{{{0k}},{{2}}}^0$  are defined in (\ref{eq:gamma_k_bar_A01}) and (\ref{eq:gamma_k_bar_A03}), respectively.
 \begin{figure*}[!ht]
\begin{align}
& \frac{1}{{{N_t}}}{\left( {{\bf{h}}_{0k}^0} \right)^H}{\left( {{\bf{V}}_{eq}^0} \right)^H}{{{\bf{\hat h}}}_{eq,0k}}{\left( {{{{\bf{\hat h}}}_{eq,0k}}} \right)^H}{\bf{V}}_{eq}^0{\bf{h}}_{0k}^0 \nonumber \\
  & = \frac{1}{{{N_t}}}{\left( {{\bf{h}}_{0k}^0} \right)^H}{\left( {{\bf{V}}_{eq}^0} \right)^H}\sqrt {{P_0}} {\bf{V}}_{eq}^0{\bf{R}}_{0k}^0{\left( {{\bf{V}}_{eq}^0} \right)^H} {\left( {{N_0}{{\bf{I}}_K} + \tau {P_0}{\bf{V}}_{eq}^0{\bf{R}}_{0k}^0{{\left( {{\bf{V}}_{eq}^0} \right)}^H}} \right)^{ - 1}}  \left( {\sqrt {{P_0}} \tau {\bf{V}}_{eq}^0{\bf{h}}_{0k}^0 + {\bf{V}}_{eq}^0{{{\bf{\tilde n}}}_{eq}}} \right) \nonumber \\
& \times  \sqrt {{P_0}} {\left( {\sqrt {{P_0}} \tau {\bf{V}}_{eq}^0{\bf{h}}_{0k}^0 + {\bf{V}}_{eq}^0{{{\bf{\tilde n}}}_{eq}}} \right)^H} {\left( {{N_0}{{\bf{I}}_K} + \tau {P_0}{\bf{V}}_{eq}^0{\bf{R}}_{0k}^0{{\left( {{\bf{V}}_{eq}^0} \right)}^H}} \right)^{ - 1}}{\bf{V}}_{eq}^0{\bf{R}}_{0k}^0
 {\left( {{\bf{V}}_{eq}^0} \right)^H}{\bf{V}}_{eq}^0{\bf{h}}_{0k}^0 \nonumber \\
& {{ = }}\frac{1}{{{N_t}}}{P_0}{\left( {\sqrt {{P_0}} \tau {\bf{V}}_{eq}^0{\bf{h}}_{0k}^0 + {\bf{V}}_{eq}^0{{{\bf{\tilde n}}}_{eq}}} \right)^H}{\left( {{N_0}{{\bf{I}}_K} + \tau {P_0}{\bf{V}}_{eq}^0{\bf{R}}_{0k}^0{{\left( {{\bf{V}}_{eq}^0} \right)}^H}} \right)^{ - 1}}{\bf{V}}_{eq}^0{\bf{R}}_{0k}^0{\left( {{\bf{V}}_{eq}^0} \right)^H}{\bf{V}}_{eq}^0{\bf{h}}_{0k}^0  \nonumber \\
& \times {\left( {{\bf{h}}_{0k}^0} \right)^H}{\left( {{\bf{V}}_{eq}^0} \right)^H}{\bf{V}}_{eq}^0{\bf{R}}_{0k}^0{\left( {{\bf{V}}_{eq}^0} \right)^H}{\left( {{N_0}{{\bf{I}}_K} + \tau {P_0}{\bf{V}}_{eq}^0{\bf{R}}_{0k}^0{{\left( {{\bf{V}}_{eq}^0} \right)}^H}} \right)^{ - 1}}\left( {\sqrt {{P_0}} \tau {\bf{V}}_{eq}^0{\bf{h}}_{0k}^0 + {\bf{V}}_{eq}^0{{{\bf{\tilde n}}}_{eq}}} \right) \nonumber \\
& \mathop  \to \limits^{{N_t} \to \infty } \frac{1}{{{N_t}}}{P_0}{\left( {\sqrt {{P_0}} \tau {\bf{V}}_{eq}^0{\bf{h}}_{0k}^0 + {\bf{V}}_{eq}^0{{{\bf{\tilde n}}}_{eq}}} \right)^H}{\left( {{N_0}{{\bf{I}}_K} + \tau {P_0}{\bf{A}}_{{{0 k}},{{1}}}^0} \right)^{ - 1}}{\bf{A}}_{{{0k}},{{1}}}^0{\bf{V}}_{eq}^0{\bf{h}}_{0k}^0 \nonumber \\
& \hspace{1cm} \times {\left( {{\bf{h}}_{0k}^0} \right)^H}{\left( {{\bf{V}}_{eq}^0} \right)^H}{\bf{A}}_{{{0k}},{{1}}}^0{\left( {{N_0}{{\bf{I}}_K} + \tau {P_0}{\bf{A}}_{{{0k}},{{1}}}^0} \right)^{ - 1}}\left( {\sqrt {{P_0}} \tau {\bf{V}}_{eq}^0{\bf{h}}_{0k}^0 + {\bf{V}}_{eq}^0{{{\bf{\tilde n}}}_{eq}}} \right) \nonumber \\
 &  = \frac{1}{{{N_t}}}{P_0}\left( {\sqrt {{P_0}} \tau {{\left( {{\bf{h}}_{0k}^0} \right)}^H}{{\left( {{\bf{V}}_{eq}^0} \right)}^H}{\bf{A}}_{{{0k}},{{2}}}^0{\bf{A}}_{{{0k}},{{1}}}^0{\bf{V}}_{eq}^0{\bf{h}}_{0k}^0
 + {{\left( {{\bf{{\tilde{n}}}}_{eq}} \right)}^H}{{\left( {{\bf{V}}_{eq}^0} \right)}^H}{\bf{A}}_{{{0k}},{{2}}}^0 {\bf{A}}_{{{0k}},{{1}}}^0{\bf{V}}_{eq}^0{\bf{h}}_{0k}^0} \right) \nonumber \\
 & \hspace{2cm} \times \left( {\sqrt {{P_0}} \tau {{\left( {{\bf{h}}_{0k}^0} \right)}^H}{{\left( {{\bf{V}}_{eq}^0} \right)}^H}{\bf{A}}_{{{0k}},{{1}}}^0{\bf{A}}_{{{0k}},{{2}}}^0{\bf{V}}_{eq}^0{\bf{h}}_{0k}^0{{ + }}{{\left( {{\bf{h}}_{0k}^0} \right)}^H}{{\left( {{\bf{V}}_{eq}^0} \right)}^H}{\bf{A}}_{{{0k}},{{1}}}^0{\bf{A}}_{{{0k}},{{2}}}^0{\bf{V}}_{eq}^0 {{\bf{{\tilde{n}}}}_{eq}} } \right), \label{eq:h0k_v_eq_1}
\end{align}
\vspace*{6pt}

 \hrulefill
\vspace*{4pt}
\end{figure*}

According to the definition of ${{\bf{V}}_{eq}^0}$ in (\ref{V0}), we obtain
\begin{align}
& \frac{1}{{{N_t}}}{\left( {{\bf{h}}_{0k}^0} \right)^H}{\left( {{\bf{V}}_{eq}^0} \right)^H}{\bf{A}}_{{{0}}k,{{2}}}^0{\bf{A}}_{{{0}}k,{{1}}}^0{\bf{V}}_{eq}^0{\bf{h}}_{0k}^0 \nonumber \\
& {{ = }}\frac{1}{{{N_t}}}{\left( {{\bf{h}}_{0k}^0} \right)^H}\left[ {{\bf{h}}_{01}^0, \cdots {\bf{h}}_{0K}^0} \right]{\bf{R}}_{\rm{0}}^{{\rm{ - 1/2}}}{\bf{A}}_{{{0}}k,{{2}}}^0{\bf{A}}_{{{0}}k,{{1}}}^0 \nonumber \\
& \hspace{2cm} \times {\bf{R}}_{{0}}^{{{ - 1/2}}}{\left[ {{\bf{h}}_{01}^0, \cdots {\bf{h}}_{0K}^0} \right]^H}{\bf{h}}_{0k}^0 \nonumber \\
& {{ = }}\frac{1}{{{N_t}}}\sum\limits_{t = 1}^K {{{\left( {{\bf{h}}_{0k}^0} \right)}^H}{\bf{h}}_{0t}^0} {\left[ {{\bf{A}}_{{{0}}k,{{5}}}^0} \right]_{tt}}{\left( {{\bf{h}}_{0t}^0} \right)^H}{\bf{h}}_{0k}^0. \label{eq:h0_v01}
\end{align}

We can further simplify (\ref{eq:h0_v01}) as follows:
\begin{align}
& \frac{1}{{{N_t}}}{\left[ {{\bf{A}}_{{{0}}k,{{5}}}^0} \right]_{tt}}{\left( {{\bf{h}}_{0k}^0} \right)^H}{\bf{h}}_{0t}^0{\left( {{\bf{h}}_{0t}^0} \right)^H}{\bf{h}}_{0k}^0 \nonumber \\
& \mathop  \to \limits^{{N_t} \to \infty } \frac{1}{{{N_t}}}{\left[ {{\bf{A}}_{{{0}}k,{{5}}}^0} \right]_{tt}}{\rm tr}\left( {{\bf{h}}_{0t}^0{{\left( {{\bf{h}}_{0t}^0} \right)}^H}{\bf{R}}_{0k}^0} \right) \nonumber \\
& \mathop  \to \limits^{{N_t} \to \infty } \frac{1}{{{N_t}}}{\left[ {{\bf{A}}_{{{0}}k,{{5}}}^0} \right]_{tt}}{\rm tr}\left( {{\bf{R}}_{0k}^0{\bf{R}}_{0t}^0} \right) \nonumber \\
&\frac{1}{{{N_t}}}{\left[ {{\bf{A}}_{{{0}}k,{{5}}}^0} \right]_{kk}}{\left( {{\bf{h}}_{0k}^0} \right)^H}{\bf{h}}_{0k}^0{\left( {{\bf{h}}_{0k}^0} \right)^H}{\bf{h}}_{0k}^0  \nonumber \\
& \mathop  \to \limits^{{N_t} \to \infty } \frac{1}{{{N_t}}}{\left[ {{\bf{A}}_{{{0}}k,{{5}}}^0} \right]_{kk}}{\rm{t}}{{\rm{r}}^{\rm{2}}}\left( {{\bf{R}}_{0k}^0} \right).
\end{align}

Then, we have
\begin{align}
&  \frac{1}{{{N_t}}}{\left( {{\bf{h}}_{0k}^0} \right)^H}{\left( {{\bf{V}}_{eq}^0} \right)^H}{\bf{A}}_{{{0}}k,{{2}}}^0{\bf{A}}_{{{0}}k,{{1}}}^0{\bf{V}}_{eq}^0{\bf{h}}_{0k}^0 \nonumber \\
& \mathop  \to \limits^{{N_t} \to \infty } \frac{1}{{{N_t}}}{\left[ {{\bf{A}}_{{{0}}k,{{5}}}^0} \right]_{kk}}{\rm tr}^{{2}}\left( {{\bf{R}}_{0k}^0} \right) \nonumber \\
&{{ + }}\frac{1}{{{N_t}}}\sum\limits_{t = 1,t \ne k}^K {{{\left[ {{\bf{A}}_{{{0}}k,{{5}}}^0} \right]}_{tt}}{\rm tr}\left( {{\bf{R}}_{0k}^0{\bf{R}}_{0t}^0} \right)}. \label{eq:h0_v0_2}
\end{align}

Also, we have
\begin{align}
& \frac{1}{{{N_t}}}{\left( {{{\bf{\tilde{n}}}_{eq}}} \right)^H}{\left( {{\bf{V}}_{eq}^0} \right)^H}  {\bf{A}}_{{{0}}k,{{2}}}^0{\bf{A}}_{{{0}}k,{{1}}}^0{\bf{V}}_{eq}^0{\bf{h}}_{0k}^0{\left( {{\bf{h}}_{0k}^0} \right)^H}{\left( {{\bf{V}}_{eq}^0} \right)^H} \nonumber \\
& \hspace{3cm} \times {\bf{A}}_{{{0}}k,{{1}}}^0 {\bf{A}}_{{{0}}k,{{2}}}^0{\bf{V}}_{eq}^0{{\bf{\tilde{n}}}_{eq}} \nonumber \\
& \mathop  \to \limits^{{N_t} \to \infty } \frac{1}{{{N_t}}}\tau {N_0}{\rm{tr}}\left({{\left( {{\bf{V}}_{eq}^0} \right)}^H}{\bf{A}}_{{{0}}k,{{2}}}^0{\bf{A}}_{{{0}}k,{{1}}}^0{\bf{V}}_{eq}^0{\bf{h}}_{0k}^0{{\left( {{\bf{h}}_{0k}^0} \right)}^H}{{\left( {{\bf{V}}_{eq}^0} \right)}^H} \right. \nonumber \\
& \hspace{3cm} \times \left.{\bf{A}}_{{{0}}k,{{1}}}^0{\bf{A}}_{{{0}}k,{{2}}}^0{\bf{V}}_{eq}^0 \right) \nonumber \\
& {{ = }}\frac{1}{{{N_t}}}\tau {N_0}{\left( {{\bf{h}}_{0k}^0} \right)^H}{\left( {{\bf{V}}_{eq}^0} \right)^H}{\left( {{\bf{A}}_{{{0}}k,{{1}}}^0{\bf{A}}_{{{0}}k,{{2}}}^0} \right)^{{2}}}{\bf{V}}_{eq}^0{\bf{h}}_{0k}^0 \nonumber \\
& {{ = }}\frac{1}{{{N_t}}}\tau {N_0}{\left( {{\bf{h}}_{0k}^0} \right)^H}\left[ {{\bf{h}}_{01}^0, \cdots {\bf{h}}_{0K}^0} \right]{\left( {{\bf{A}}_{{{0}}k,{{1}}}^0{\bf{A}}_{{{0}}k,{{2}}}^0} \right)^{{2}}} \nonumber \\
& \hspace{3cm} \times {\bf{R}}_{{0}}^{{{ - 1}}}{\left[ {{\bf{h}}_{01}^0, \cdots {\bf{h}}_{0K}^0} \right]^H}{\bf{h}}_{0k}^0 \nonumber \\
& \mathop  \to \limits^{{N_t} \to \infty } \frac{1}{{{N_t}}}\tau {N_0}\left( {{\left[ {{\bf{A}}_{{{0}}k,{{4}}}^0} \right]}_{kk}}{\rm{t}}{{\rm{r}}^2}\left( {{\bf{R}}_{0k}^0} \right) \right. \nonumber \\
& \hspace{3cm} \left. + \sum\limits_{t = 1,t \ne k}^K {{{\left[ {{\bf{A}}_{{{0}}k,{{4}}}^0} \right]}_{tt}}{\rm{tr}}\left( {{\bf{R}}_{0k}^0{\bf{R}}_{0t}^0} \right)}  \right).  \label{eq:h0_w0}
\end{align}

Substituting (\ref{eq:h0_v0_2}) and (\ref{eq:h0_w0}) into (\ref{eq:h0k_v_eq_1}), we obtain
\begin{align}
&\frac{1}{{{N_t}}}{\left( {{\bf{h}}_{0k}^0} \right)^H}{\left( {{\bf{V}}_{eq}^0} \right)^H}{{{\bf{\hat h}}}_{eq,0k}}{\left( {{{{\bf{\hat h}}}_{eq,0k}}} \right)^H}{\bf{V}}_{eq}^0{\bf{h}}_{0k}^0 \nonumber \\
& \mathop  \to \limits^{{N_t} \to \infty } \frac{1}{{{N_t}}}{P_0}\left( {{P_0}{\tau ^{{2}}}{{\left( {{{\left[ {{\bf{A}}_{{{0k}},{{5}}}^0} \right]}_{kk}}{{\rm tr}^{{2}}}\left( {{\bf{R}}_{0k}^0} \right)} \right. } } } \right. \\
& \left. {{{\left.{+ \sum\limits_{t = 1,t \ne k}^K {{{\left[ {{\bf{A}}_{{{0k}},{{5}}}^0} \right]}_{tt}}{\rm tr}\left( {{\bf{R}}_{0k}^0{\bf{R}}_{0t}^0} \right)} } \right)}^{\rm{2}}}} \right) {{ + }}\frac{1}{{{N_t}}}{P_0}{N_0}\tau  \nonumber \\
 & \times \left( {{{\left[ {{\bf{A}}_{{{0}}k,{{4}}}^0} \right]}_{kk}}{{\rm tr}^2}\left( {{\bf{R}}_{0k}^0} \right) + \sum\limits_{t = 1,t \ne k}^K {{{\left[ {{\bf{A}}_{{{0k}},{{4}}}^0} \right]}_{tt}}{\rm tr} \left( {{\bf{R}}_{0k}^0{\bf{R}}_{0t}^0} \right)} } \right) \nonumber \\
&  = a_{{{0}}k,2}^0. \label{eq:h0k_v_eq_2}
\end{align}

Substituting (\ref{eq:h0_v0_2}) and (\ref{eq:h0k_v_eq_2}) into (\ref{eq:hk_est}), we obtain
\begin{align}
\frac{1}{{{N_t}}}{\left| {g_{0k,k}^0} \right|^2}\mathop  \to \limits^{{N_t} \to \infty } \frac{P}{{{N_t}}}\frac{{a_{0k,2}^0}}{{a_{0k,1}^0}}. \label{eq:g0kk2}
\end{align}

For ${\left| {g_{0t,k}^0} \right|^2}$, we obtain
\begin{align}
{\left| {g_{0t,k}^0} \right|^2} = P{\left( {{\bf{h}}_{0k}^0} \right)^H}  \left( {{\bf{V}}_{eq}^0} \right)^H  \frac{{{{{\bf{\hat h}}}_{eq,0t}}{\bf{\hat h}}{{_{eq,0t}^H}_{}}_{}}}
{{{{\left\| {{{{\bf{\hat h}}}_{eq,0t}}} \right\|}^2}}}{{\bf{V}}_{eq}^0}{\bf{h}}_{0k}^0. \label{eq:g0tk}
\end{align}

Following a similar approach as  in (\ref{eq:h_est_norm_2})--(\ref{eq:hk_est}), the denominator of (\ref{eq:g0tk})
is given by
\begin{align}
&  \frac{1}{{{N_t}}}{\left\| {{{{\bf{\hat h}}}_{eq,0t}}} \right\|^2} \mathop  \to \limits^{{N_t} \to \infty } \nonumber \\
& \frac{1}{{{N_t}}}{P_0}\left[ {{P_0}{\tau ^{\rm{2}}}\left( {\sum\limits_{p = 1,p \ne t}^K {{{\left[ {{\bf{A}}_{0t,{{4}}}^0} \right]}_{pp}}{\rm{tr}}\left( {{\bf{R}}_{0t}^0{\bf{R}}_{0p}^0} \right)   } } \right. } \right.  \nonumber \\
& \left.{ \left. { { + {{\left[ {{\bf{A}}_{0t,{{4}}}^0} \right]}_{tt}}{\rm{t}}{{\rm{r}}^2}\left( {{\bf{R}}_{0t}^0} \right)} } \right) + \tau {N_0}\sum\limits_{p = 1}^K {{{\left[ {{\bf{A}}_{0t,{{4}}}^0} \right]}_{pp}}{\rm{tr}}\left( {{\bf{R}}_{0p}^0} \right)} } \! \right] = a_{{{0}}t,{{1}}}^0. \label{eq:h0t_est}
\end{align}

For the numerator of (\ref{eq:g0tk}), we have (\ref{eq:hk_ht_veq}) given at the top of the next page.
 \begin{figure*}[!ht]
\begin{align}
& \frac{1}{{{N_t}}}{\left( {{\bf{h}}_{0k}^0} \right)^H}{\left( {{\bf{V}}_{eq}^0} \right)^H}{{{\bf{\hat h}}}_{eq,0t}} {\bf{\hat h}}_{{eq,0t}}^H{\bf{V}}_{eq}^0
{\bf{h}}_{0k}^0 \nonumber \\
& = \frac{1}{{{N_t}}}{\left( {{\bf{h}}_{0k}^0} \right)^H}{\left( {{\bf{V}}_{eq}^0} \right)^H}\sqrt {{P_0}} {\bf{V}}_{eq}^0{\bf{R}}_{0t}^0{\left( {{\bf{V}}_{eq}^0} \right)^H}{\left( {{N_0}{{\bf{I}}_K} + \tau {P_0}{\bf{V}}_{eq}^0{\bf{R}}_{0t}^0{{\left( {{\bf{V}}_{eq}^0} \right)}^H}} \right)^{ - 1}} \nonumber \\
& \times \left( {\sqrt {{P_0}} \tau {\bf{V}}_{eq}^0{\bf{h}}_{0t}^0 + {\bf{V}}_{eq}^0 {{\bf{{\tilde{n}}}}_{eq}}  } \right)\sqrt {{P_0}} {\left( {\sqrt {{P_0}} \tau {\bf{V}}_{eq}^0{\bf{h}}_{0t}^0 + {\bf{V}}_{eq}^0 {{\bf{{\tilde{n}}}}_{eq}} } \right)^H}{\left( {{N_0}{{\bf{I}}_K} + \tau {P_0}{\bf{V}}_{eq}^0{\bf{R}}_{0t}^0{{\left( {{\bf{V}}_{eq}^0} \right)}^H}} \right)^{ - 1}} \nonumber \\
&  \times {\bf{V}}_{eq}^0{\bf{R}}_{0t}^0{\left( {{\bf{V}}_{eq}^0} \right)^H} {\bf{V}}_{eq}^0{\bf{h}}_{0k}^0 \nonumber \\
&  \mathop  \to \limits^{{N_t} \to \infty } \frac{1}{{{N_t}}}{P_0}{\left( {{\bf{h}}_{0k}^0} \right)^H}{\left( {{\bf{V}}_{eq}^0} \right)^H}{\bf{A}}_{{{0}}t,{{1}}}^0{\bf{A}}_{{{0}}t,{{2}}}^0\left( {\sqrt {{P_0}} \tau {\bf{V}}_{eq}^0{\bf{h}}_{0t}^0 + {\bf{V}}_{eq}^0 {{\bf{{\tilde{n}}}}_{eq}} } \right) \nonumber \\
&  \times {\left( {\sqrt {{P_0}} \tau {\bf{V}}_{eq}^0{\bf{h}}_{0t}^0 + {\bf{V}}_{eq}^0 {{\bf{{\tilde{n}}}}_{eq}}} \right)^H}{\bf{A}}_{{{0}}t,{{2}}}^0{\bf{A}}_{{{0}}t,{{1}}}^0{\bf{V}}_{eq}^0{\bf{h}}_{0k}^0 \nonumber \\
&   {{ = }}\frac{1}{{{N_t}}}{\left( {{P_0}\tau } \right)^{\rm{2}}}{\left( {{\bf{h}}_{0k}^0} \right)^H}{\left( {{\bf{V}}_{eq}^0} \right)^H}{\bf{A}}_{{{0}}t,{{1}}}^0{\bf{A}}_{{{0}}t,{{2}}}^0{\bf{V}}_{eq}^0{\bf{h}}_{0t}^0{\left( {{\bf{V}}_{eq}^0{\bf{h}}_{0t}^0} \right)^H}{\bf{A}}_{{{0}}t,{{2}}}^0{\bf{A}}_{{{0}}t,{{1}}}^0{\bf{V}}_{eq}^0{\bf{h}}_{0k}^0 \nonumber \\
&  {{ + }}\frac{1}{{{N_t}}}{P_0}{\left( {{\bf{h}}_{0k}^0} \right)^H}{\left( {{\bf{V}}_{eq}^0} \right)^H}{\bf{A}}_{{{0}}t,{{1}}}^0{\bf{A}}_{{{0}}t,{{2}}}^0{\bf{V}}_{eq}^0 {{\bf{{\tilde{n}}}}_{eq}}{\left( {{\bf{V}}_{eq}^0 {{\bf{{\tilde{n}}}}_{eq}} } \right)^H}{\bf{A}}_{{{0}}t,{{2}}}^0{\bf{A}}_{{{0}}t,{{1}}}^0{\bf{V}}_{eq}^0{\bf{h}}_{0k}^0. \label{eq:hk_ht_veq}
\end{align}
\vspace*{6pt}

 \hrulefill
\vspace*{4pt}
\end{figure*}

Substituting the expression for ${{\bf{V}}_{eq}^0}$ in (\ref{V0}) into (\ref{eq:hk_ht_veq}), we obtain (\ref{eq:hk_ht_veq_2}) given
at the top of the next page.
 \begin{figure*}[!ht]
\begin{align}
& \frac{1}{{{N_t}}}{\left( {{\bf{h}}_{0k}^0} \right)^H}{\left( {{\bf{V}}_{eq}^0} \right)^H}{{{\bf{\hat h}}}_{eq,0t}} {\bf{\hat h}}_{{eq,0t}}^H{\bf{V}}_{eq}^0
{\bf{h}}_{0k}^0 \nonumber \\
& {{ = }}\frac{1}{{{N_t}}}{\left( {{P_0}\tau } \right)^{\rm{2}}}{\left( {{\bf{h}}_{0k}^0} \right)^H}\left[ {{\bf{h}}_{01}^0, \cdots {\bf{h}}_{0K}^0} \right]{\bf{R}}_0^{ - 1/2}{\bf{A}}_{{{0}}t,{{1}}}^0{\bf{A}}_{{{0}}t,{{2}}}^0{\bf{R}}_0^{ - 1/2}{\left[ {{\bf{h}}_{01}^0, \cdots {\bf{h}}_{0K}^0} \right]^H}{\bf{h}}_{0t}^0 \nonumber \\
& \times {\left( {{\bf{h}}_{0t}^0} \right)^H}{\left[ {{\bf{h}}_{01}^0, \cdots {\bf{h}}_{0K}^0} \right]^H}{\bf{R}}_0^{ - 1/2}{\bf{A}}_{{{0}}t,{{2}}}^0{\bf{A}}_{{{0}}t,{{1}}}^0{\bf{R}}_0^{ - 1/2}{\left[ {{\bf{h}}_{01}^0, \cdots {\bf{h}}_{0K}^0} \right]^H}{\bf{h}}_{0k}^0 \nonumber \\
& {{ + }}\frac{1}{{{N_t}}}{P_0}{\left( {{\bf{h}}_{0k}^0} \right)^H}\left[ {{\bf{h}}_{01}^0, \cdots {\bf{h}}_{0K}^0} \right]{\bf{R}}_0^{ - 1/2}{\bf{A}}_{{{0}}t,{{1}}}^0{\bf{A}}_{{{0}}t,{{2}}}^0{\bf{R}}_0^{ - 1/2}{\left[ {{\bf{h}}_{01}^0, \cdots {\bf{h}}_{0K}^0} \right]^H} {{\bf{{\tilde{n}}}}_{eq}} \nonumber \\
& \times {\left( {{\bf{{\tilde{n}}}}_{eq}} \right)^H}\left[ {{\bf{h}}_{01}^0, \cdots {\bf{h}}_{0K}^0} \right]{\bf{R}}_0^{ - 1/2}{\bf{A}}_{{{0}}t,{{2}}}^0{\bf{A}}_{{{0}}t,{{1}}}^0{\bf{R}}_0^{ - 1/2}\left[ {{\bf{h}}_{01}^0, \cdots {\bf{h}}_{0K}^0} \right]{\bf{h}}_{0k}^0 \nonumber \\
& {{ = }}\frac{1}{{{N_t}}}{\left( {{P_0}\tau } \right)^{{2}}}{\left( {{\bf{h}}_{0k}^0} \right)^H}\left[ {{\bf{h}}_{01}^0, \cdots {\bf{h}}_{0K}^0} \right]
{\bf{A}}_{{{0}}t,{{5}}}^0 {\left[ {{\bf{h}}_{01}^0, \cdots {\bf{h}}_{0K}^0} \right]^H}{\bf{h}}_{0t}^0 \nonumber \\
& \times {\left( {{\bf{h}}_{0t}^0} \right)^H}{\left[ {{\bf{h}}_{01}^0, \cdots {\bf{h}}_{0K}^0} \right]^H} {\bf{A}}_{{{0}}t,{{5}}}^0 {\left[ {{\bf{h}}_{01}^0, \cdots {\bf{h}}_{0K}^0} \right]^H}{\bf{h}}_{0k}^0 \nonumber \\
& {{ + }}\frac{1}{{{N_t}}}{P_0}{\left( {{\bf{h}}_{0k}^0} \right)^H}\left[ {{\bf{h}}_{01}^0, \cdots {\bf{h}}_{0K}^0} \right] {\bf{A}}_{{{0}}t,{{5}}}^0 {\left[ {{\bf{h}}_{01}^0, \cdots {\bf{h}}_{0K}^0} \right]^H} {{\bf{{\tilde{n}}}}_{eq}} \nonumber \\
& \times {\left( {{\bf{{\tilde{n}}}}_{eq}} \right)^H}\left[ {{\bf{h}}_{01}^0, \cdots {\bf{h}}_{0K}^0} \right] {\bf{A}}_{{{0}}t,{{5}}}^0 \left[ {{\bf{h}}_{01}^0, \cdots {\bf{h}}_{0K}^0} \right]{\bf{h}}_{0k}^0 . \label{eq:hk_ht_veq_2}
\end{align}

\vspace*{6pt}

 \hrulefill
\vspace*{4pt}
\end{figure*}

By applying \cite[Corollary 1]{Evans2000TIT} to ${\left( {{\bf{h}}_{0k}^0} \right)^H}\left[ {{\bf{h}}_{01}^0, \cdots {\bf{h}}_{0K}^0} \right]$,
we have
\begin{align}
& \frac{1}{{{N_t}}}{\left( {{\bf{h}}_{0k}^0} \right)^H}{\left( {{\bf{V}}_{eq}^0} \right)^H}{{{\bf{\hat h}}}_{eq,0t}} {\bf{\hat h}}_{{eq,0t}}^H{\bf{V}}_{eq}^0
{\bf{h}}_{0k}^0 \nonumber \\
& \mathop  \to \limits^{{N_t} \to \infty } \frac{1}{{{N_t}}}{\left( {{P_0}\tau } \right)^{{2}}}{\rm tr}\left( {{\bf{R}}_{0k}^0} \right){\left[ {{\bf{A}}_{0t,{{5}}}^0} \right]_{kk}}{\bf{e}}_k^H{\left[ {{\bf{h}}_{01}^0, \cdots {\bf{h}}_{0K}^0} \right]^H}{\bf{h}}_{0t}^0 \nonumber \\
& \times {\left( {{\bf{h}}_{0t}^0} \right)^H}\left[ {{\bf{h}}_{01}^0, \cdots {\bf{h}}_{0K}^0} \right]{{\bf{e}}_k}{\rm tr}\left( {{\bf{R}}_{0k}^0} \right){\left[ {{\bf{A}}_{0t,{{5}}}^0} \right]_{kk}}
\nonumber \\
 & {{ + }}\frac{1}{{{N_t}}}{P_0}{\rm tr}\left( {{\bf{R}}_{0k}^0} \right){\left[ {{\bf{A}}_{{{0}}t,{{5}}}^0} \right]_{kk}}{\bf{e}}_k^H{\left[ {{\bf{h}}_{01}^0, \cdots {\bf{h}}_{0K}^0} \right]^H} {{\bf{{\tilde{n}}}}_{eq}} \nonumber \\
& \times {\left( {{\bf{{\tilde{n}}}}_{eq}} \right)^H}\left[ {{\bf{h}}_{01}^0, \cdots {\bf{h}}_{0K}^0} \right]_0^{ - 1/2}{{\bf{e}}_k}{\rm tr}\left( {{\bf{R}}_{0k}^0} \right){\left[ {{\bf{A}}_{0t,{{5}}}^0} \right]_{kk}},
 \end{align}
which can be simplified further as
\begin{align}
& \frac{1}{{{N_t}}}{\left( {{\bf{h}}_{0k}^0} \right)^H}{\left( {{\bf{V}}_{eq}^0} \right)^H}{{{\bf{\hat h}}}_{eq,0t}} {\bf{\hat h}}_{{eq,0t}}^H{\bf{V}}_{eq}^0
{\bf{h}}_{0k}^0 \nonumber \\
& {{ = }}\frac{1}{{{N_t}}}{\left( {{P_0}\tau {\rm tr} \left( {{\bf{R}}_{0k}^0} \right){{\left[ {{\bf{A}}_{{{0t}},{{5}}}^0} \right]}_{kk}}} \right)^{{2}}}{\left( {{\bf{h}}_{0k}^0} \right)^H}{\bf{h}}_{0t}^0{\left( {{\bf{h}}_{0t}^0} \right)^H}{\bf{h}}_{0k}^0 \nonumber \\
& + \frac{1}{{{N_t}}}{P_0}{\left( {{\rm tr}\left( {{\bf{R}}_{0k}^0} \right){{\left[ {{\bf{A}}_{{{0}}t,{{5}}}^0} \right]}_{kk}}} \right)^{{2}}}{\left( {{\bf{h}}_{0k}^0} \right)^H}{{\bf{{\tilde{n}}}}_{eq}} {\left( {{\bf{{\tilde{n}}}}_{eq}} \right)^H}{\bf{h}}_{0k}^0 \nonumber \\
& \mathop  \to \limits^{{N_t} \to \infty } \frac{1}{{{N_t}}}{\left( {{P_0}\tau {\rm tr}\left( {{\bf{R}}_{0k}^0} \right){{\left[ {{\bf{A}}_{{{0}}t,{{5}}}^0} \right]}_{kk}}} \right)^{{2}}}{\rm tr}\left( {{\bf{R}}_{0k}^0{\bf{R}}_{0t}^0} \right) \nonumber \\
&  \hspace{2cm} + \frac{1}{{{N_t}}}{P_0}\tau {N_0}{\rm tr}{^{{3}}}\left( {{\bf{R}}_{0k}^0} \right){\left( {{{\left[ {{\bf{A}}_{{{0}}t,{{5}}}^0} \right]}_{kk}}} \right)^{{2}}} \nonumber \\
& {{ = }}b_{{{0}}t,{{k}}}^0.  \label{eq:hk_ht_veq_3}
\end{align}

By combining (\ref{eq:g0kk2}), (\ref{eq:h0t_est}), and (\ref{eq:hk_ht_veq_3}), we obtain
\begin{align}
\frac{1}{{{N_t}}}{\left| {g_{0t,k}^0} \right|^2}\mathop  \to \limits^{{N_t} \to \infty } \frac{P}{{{N_t}}}\frac{{b_{{{0}}t,k}^0}}{{a_{{{0}}t,{{1}}}^0}}. \label{eq:g0tk}
\end{align}

Then, we consider ${\left| {g_{lt,k}^0} \right|^2}$, which leads to
\begin{align}
\frac{P}{{{N_t}}}{\left| {g_{lt,k}^0} \right|^2}{{ = }}\frac{P}{{{N_t}}}{\left( {{\bf{h}}_{lk}^0} \right)^H} \left( {{\bf{V}}_{eq}^l} \right)^H \frac{{{{{\bf{\hat h}}}_{eq,lt}}{{\left( {{{{\bf{\hat h}}}_{eq,lt}}} \right)}^H}}}{{{{\left\| {{{{\bf{\hat h}}}_{eq,lt}}} \right\|}^2}}} {{\bf{V}}_{eq}^l} {\bf{h}}_{lk}^0. \label{eq:gltk}
\end{align}

Following a similar approach as in (\ref{eq:h_est_norm_2})--(\ref{eq:hk_est}), the denominator of (\ref{eq:gltk})
is obtained as
\begin{align}
  & \frac{1}{{{N_t}}}{\left\| {{{{\bf{\hat h}}}_{eq,lt}}} \right\|^2} \mathop  \to \limits^{{N_t} \to \infty } \frac{1}{{{N_t}}}{P_I} \times \nonumber \\
& \left[{P_I}{\tau ^{{2}}} \left[ {\sum\limits_{p = 1,p \ne t}^K {{{\left[ {{\bf{A}}_{lt,{{4}}}^l} \right]}_{pp}}{\rm tr} \left( {{\bf{R}}_{lt}^l{\bf{R}}_{lp}^l} \right) + {{\left[ {{\bf{A}}_{lt,{{4}}}^l} \right]}_{tt}}{\rm tr}^2\left( {{\bf{R}}_{lt}^l} \right)} } \right] \right. \nonumber \\
&  \left.{{ + }}\tau {N_0}\sum\limits_{p = 1}^K {{{\left[ {{\bf{A}}_{lt,{{4}}}^l} \right]}_{pp}}{\rm tr}\left( {{\bf{R}}_{lp}^l} \right)} \right] \nonumber\\
& {{ = }}a_{lt,{{1}}}^l.  \label{eq:alt}
\end{align}

For the numerator of (\ref{eq:gltk}), we have (\ref{eq:hlk_v_eq}) given at the top of the next page.
 \begin{figure*}[!ht]
\begin{align}
& \frac{1}{{{N_t}}}{\left( {{\bf{h}}_{lk}^0} \right)^H}{\left( {{\bf{V}}_{eq}^l} \right)^H}{{{\bf{\hat h}}}_{eq,lt}}{\bf{\hat h}}{_{eq,lt}^H}{\bf{V}}_{eq}^l{\bf{h}}_{lk}^0 \nonumber \\
 & = \frac{1}{{{N_t}}}{\left( {{\bf{h}}_{lk}^0} \right)^H}{\left( {{\bf{V}}_{eq}^l} \right)^H}\sqrt {{P_I}} {\bf{V}}_{eq}^l{\bf{R}}_{lt}^l{\left( {{\bf{V}}_{eq}^l} \right)^H}{\left( {{N_0}{{\bf{I}}_K} + \tau {P_I}{\bf{V}}_{eq}^l{\bf{R}}_{lt}^l{{\left( {{\bf{V}}_{eq}^l} \right)}^H}} \right)^{ - 1}}\left( {\sqrt {{P_I}} \tau {\bf{V}}_{eq}^l{\bf{h}}_{lt}^l + {\bf{V}}_{eq}^l {{\bf{{\tilde{n}}}}_{eq}}  } \right) \nonumber \\
& \times \sqrt {{P_I}} {\left( {\sqrt {{P_I}} \tau {\bf{V}}_{eq}^l{\bf{h}}_{lt}^l + {\bf{V}}_{eq}^l {{\bf{{\tilde{n}}}}_{eq}}} \right)^H}{\left( {{N_0}{{\bf{I}}_K} + \tau {P_I}{\bf{V}}_{eq}^l{\bf{R}}_{lt}^l{{\left( {{\bf{V}}_{eq}^l} \right)}^H}} \right)^{ - 1}}{\bf{V}}_{eq}^l{\bf{R}}_{lt}^l{\left( {{\bf{V}}_{eq}^l} \right)^H}{\bf{V}}_{eq}^l{\bf{h}}_{lk}^0 \nonumber \\
& \mathop  \to \limits^{{N_t} \to \infty } \frac{1}{{{N_t}}}{P_I}{\left( {{\bf{h}}_{lk}^0} \right)^H}{\left( {{\bf{V}}_{eq}^l} \right)^H}{\bf{A}}_{lt,{{1}}}^l{\bf{A}}_{lt,{{2}}}^l\left( {\sqrt {{P_I}} \tau {\bf{V}}_{eq}^l{\bf{h}}_{lt}^l + {\bf{V}}_{eq}^l {{\bf{{\tilde{n}}}}_{eq}}  } \right){\left( {\sqrt {{P_I}} \tau {\bf{V}}_{eq}^l{\bf{h}}_{lt}^l + {\bf{V}}_{eq}^l {{\bf{{\tilde{n}}}}_{eq}} } \right)^H}{\bf{A}}_{lt,{{2}}}^l{\bf{A}}_{lt,{{1}}}^l{\bf{V}}_{eq}^l{\bf{h}}_{lk}^0 \nonumber \\
& \mathop  \to \limits^{{N_t} \to \infty } \frac{1}{{{N_t}}}{P_I}{\mathop{\rm tr}\nolimits} \left( {{\bf{R}}_{lk}^0{{\left( {{\bf{V}}_{eq}^l} \right)}^H}{\bf{A}}_{lt,{{1}}}^l{\bf{A}}_{lt,{{2}}}^l\left( {\sqrt {{P_I}} \tau {\bf{V}}_{eq}^l{\bf{h}}_{lt}^l + {\bf{V}}_{eq}^l {{\bf{{\tilde{n}}}}_{eq}} } \right){{\left( {\sqrt {{P_I}} \tau {\bf{V}}_{eq}^l{\bf{h}}_{lt}^l + {\bf{V}}_{eq}^l {{\bf{{\tilde{n}}}}_{eq}}  } \right)}^H}{\bf{A}}_{lt,{{1}}}^l{\bf{A}}_{lt,{{2}}}^l{\bf{V}}_{eq}^l} \right).
\label{eq:hlk_v_eq}
\end{align}

\vspace*{6pt}

 \hrulefill
\vspace*{4pt}
\end{figure*}

Based on (\ref{Vl}), we can re-express (\ref{eq:hlk_v_eq}) as (\ref{eq:hlk_v_eq_2}) given at the top of the next page.
 \begin{figure*}[!ht]
\begin{align}
& \frac{1}{{{N_t}}}{\left( {{\bf{h}}_{lk}^0} \right)^H}{\left( {{\bf{V}}_{eq}^l} \right)^H}{{{\bf{\hat h}}}_{eq,lt}}{\bf{\hat h}}{_{eq,lt}^H}{\bf{V}}_{eq}^l{\bf{h}}_{lk}^0 \nonumber \\
&  = \frac{1}{{{N_t}}}{P_I}{\left( {\sqrt {{P_I}} \tau {\bf{V}}_{eq}^l{\bf{h}}_{lt}^l + {\bf{V}}_{eq}^l{{{\bf{\tilde n}}}_{eq}}} \right)^H}{\bf{A}}_{lt,{{1}}}^l{\bf{A}}_{lt,{{2}}}^l{\bf{R}}_l^{ - 1/2}{\left[ {{\bf{h}}_{l1}^l, \cdots {\bf{h}}_{lK}^l} \right]^H}{\bf{R}}_{lk}^0\left[ {{\bf{h}}_{l1}^l, \cdots {\bf{h}}_{lK}^l} \right]{\bf{R}}_l^{ - 1/2} \nonumber \\
& \times {\bf{A}}_{lt,{{1}}}^l{\bf{A}}_{lt,{{2}}}^l\left( {\sqrt {{P_I}} \tau {\bf{V}}_{eq}^l{\bf{h}}_{lt}^l + {\bf{V}}_{eq}^l{{{\bf{\tilde n}}}_{eq}}} \right) \nonumber \\
& \mathop  \to \limits^{{N_t} \to \infty } \frac{1}{{{N_t}}}P_I^2{\tau ^2}{\left( {{\bf{h}}_{lt}^l} \right)^H}{\left( {{\bf{V}}_{eq}^l} \right)^H}{\bf{A}}_{lt,{{1}}}^l{\bf{A}}_{lt,{{2}}}^l{\bf{R}}_l^{ - 1/2}{\left[ {{\bf{h}}_{l1}^l, \cdots {\bf{h}}_{lK}^l} \right]^H}{\bf{R}}_{lk}^0\left[ {{\bf{h}}_{l1}^l, \cdots {\bf{h}}_{lK}^l} \right]{\bf{R}}_l^{ - 1/2}{\bf{A}}_{lt,{{1}}}^l{\bf{A}}_{lt,{{2}}}^l{\bf{V}}_{eq}^l{\bf{h}}_{lt}^l \nonumber
\end{align}
\begin{align}
&  + \frac{1}{{{N_t}}}{P_I}{\left( {{{{\bf{\tilde n}}}_{eq}}} \right)^H}{\left( {{\bf{V}}_{eq}^l} \right)^H}{\bf{A}}_{lt,{{1}}}^l{\bf{A}}_{lt,{{2}}}^l{\bf{R}}_l^{ - 1/2}{\left[ {{\bf{h}}_{l1}^l, \cdots {\bf{h}}_{lK}^l} \right]^H}{\bf{R}}_{lk}^0\left[ {{\bf{h}}_{l1}^l, \cdots {\bf{h}}_{lK}^l} \right]{\bf{R}}_l^{ - 1/2}{\bf{A}}_{lt,{{1}}}^l{\bf{A}}_{lt,{{2}}}^l{\bf{V}}_{eq}^l{{{\bf{\tilde n}}}_{eq}}. \label{eq:hlk_v_eq_2}
\end{align}
\vspace*{6pt}

 \hrulefill
\vspace*{4pt}
\end{figure*}

For the first term  on the right hand side of (\ref{eq:hlk_v_eq_2}), we have (\ref{eq:hlk_v_eq_11})
given at the top of the next page.
 \begin{figure*}[!ht]
\begin{align}
& \frac{1}{{{N_t}}}{\left( {{\bf{h}}_{lt}^l} \right)^H}{\left( {{\bf{V}}_{eq}^l} \right)^H}{\bf{A}}_{lt,{{1}}}^l{\bf{A}}_{lt,{{2}}}^l{\bf{R}}_l^{ - 1/2}{\left[ {{\bf{h}}_{l1}^l, \cdots {\bf{h}}_{lK}^l} \right]^H}{\bf{R}}_{lk}^0\left[ {{\bf{h}}_{l1}^l, \cdots {\bf{h}}_{lK}^l} \right]{\bf{R}}_l^{ - 1/2}{\bf{A}}_{lt,{{2}}}^l{\bf{A}}_{lt,{{1}}}^l{\bf{V}}_{eq}^l{\bf{h}}_{lt}^l \nonumber \\
& \mathop  \to \limits^{{N_t} \to \infty } \frac{1}{{{N_t}}}{\left( {{\bf{h}}_{lt}^l} \right)^H}\left[ {{\bf{h}}_{l1}^l, \cdots {\bf{h}}_{lK}^l} \right]{\bf{R}}_l^{ - 1/2}{\bf{A}}_{lt,{{1}}}^l{\bf{A}}_{lt,{{2}}}^l
{\bf{R}}_l^{ - 1/2}{\left[ {{\bf{h}}_{l1}^l, \cdots {\bf{h}}_{lK}^l} \right]^H} \nonumber \\
& \hspace{3cm} \times {\bf{R}}_{lk}^0\left[ {{\bf{h}}_{l1}^l, \cdots {\bf{h}}_{lK}^l} \right]{\bf{R}}_l^{ - 1/2}{\bf{A}}_{lt,{{2}}}^l{\bf{A}}_{lt,{{1}}}^l{\bf{R}}_l^{ - 1/2}{\left[ {{\bf{h}}_{l1}^l, \cdots {\bf{h}}_{lK}^l} \right]^H}{\bf{h}}_{lt}^l \nonumber \\
& \mathop  \to \limits^{{N_t} \to \infty } \frac{1}{{{N_t}}}{\rm{tr}}\left( {{\bf{R}}_{lt}^l} \right)\left[ {{\bf{A}}_{lt,{{5}}}^l} \right]{\bf{e}}_t^H{\left[ {{\bf{h}}_{l1}^l, \cdots {\bf{h}}_{lK}^l} \right]^H}{\bf{R}}_{lk}^0\left[ {{\bf{h}}_{l1}^l, \cdots {\bf{h}}_{lK}^l} \right]{{\bf{e}}_t}{\rm{tr}}\left( {{\bf{R}}_{lt}^l} \right)\left[ {{\bf{A}}_{lt,{{5}}}^l} \right] \nonumber \\
& \mathop  \to \limits^{{N_t} \to \infty } \frac{1}{{{N_t}}}{\rm{t}}{{\rm{r}}^2}\left( {{\bf{R}}_{lt}^l} \right)\left[ {{\bf{A}}_{lt,{{5}}}^l} \right]_{tt}^2{\rm{tr}}\left( {{\bf{R}}_{lt}^l{\bf{R}}_{lk}^0} \right). \label{eq:hlk_v_eq_11}
\end{align}
\vspace*{6pt}

 \hrulefill
\vspace*{4pt}
\end{figure*}

For the second term  on the right hand side of  (\ref{eq:hlk_v_eq_2}), we have (\ref{eq:hlk_v_eq_12}) given at the top of
the next page.
 \begin{figure*}[!ht]
\begin{align}
& \frac{1}{{{N_t}}}{\left( {{{{\bf{\tilde n}}}_{eq}}} \right)^H}{\left( {{\bf{V}}_{eq}^l}\right)^H}{\bf{A}}_{lt,{{1}}}^l{\bf{A}}_{lt,{{2}}}^l{\bf{R}}_l^{ - 1/2}{\left[ {{\bf{h}}_{l1}^l, \cdots {\bf{h}}_{lK}^l} \right]^H}{\bf{R}}_{lk}^0\left[ {{\bf{h}}_{l1}^l, \cdots {\bf{h}}_{lK}^l} \right]{\bf{R}}_l^{ - 1/2}{\bf{A}}_{lt,{{2}}}^l{\bf{A}}_{lt,{{1}}}^l{\bf{V}}_{eq}^l{{{\bf{\tilde n}}}_{eq}} \nonumber \\
& \mathop  \to \limits^{{N_t} \to \infty } \frac{1}{{{N_t}}}\tau {N_0}{\rm{tr}}\left( {\left[ {{\bf{h}}_{l1}^l, \cdots {\bf{h}}_{lK}^l} \right]{\bf{A}}_{lt,{{5}}}^l{{\left[ {{\bf{h}}_{l1}^l, \cdots {\bf{h}}_{lK}^l} \right]}^H}{\bf{R}}_{lk}^0\left[ {{\bf{h}}_{l1}^l, \cdots {\bf{h}}_{lK}^l} \right]{\bf{A}}_{lt,{{5}}}^l{{\left[ {{\bf{h}}_{l1}^l, \cdots {\bf{h}}_{lK}^l} \right]}^H}} \right) \nonumber \\
& {{ = }}\frac{1}{{{N_t}}}\tau {N_0}{\rm{tr}}\left( {\sum\limits_{p = 1}^K {{{\left[ {{\bf{A}}_{lt,{{5}}}^l} \right]}_{pp}}{\bf{h}}_{lp}^l{{\left( {{\bf{h}}_{lp}^l} \right)}^H}{\bf{R}}_{lk}^0\sum\limits_{m = 1}^K {{{\left[ {{\bf{A}}_{lt,{{5}}}^l} \right]}_{mm}}{\bf{h}}_{lm}^l{{\left( {{\bf{h}}_{lm}^l} \right)}^H}} } } \right) \nonumber \\
& {{ = }}\frac{1}{{{N_t}}}\tau {N_0}\left( {\sum\limits_{p = 1}^K {\left[ {{\bf{A}}_{lt,{{5}}}^l} \right]_{pp}^2{\rm{tr}}\left( {{\bf{h}}_{lp}^l{{\left( {{\bf{h}}_{lp}^l} \right)}^H}{\bf{R}}_{lk}^0{\bf{h}}_{lp}^l{{\left( {{\bf{h}}_{lp}^l} \right)}^H}} \right)} } \right) \nonumber \\
& + \sum\limits_{p = 1}^K {\sum\limits_{m = 1,m \ne p}^K {{{\left[ {{\bf{A}}_{lt,{{5}}}^l} \right]}_{pp}}{{\left[ {{\bf{A}}_{lt,{{5}}}^l} \right]}_{mm}}{\rm{tr}}\left( {{\bf{h}}_{lp}^l{{\left( {{\bf{h}}_{lp}^l} \right)}^H}{\bf{R}}_{lk}^0{\bf{h}}_{lm}^l{{\left( {{\bf{h}}_{lm}^l} \right)}^H}} \right)} } \nonumber \\
& \mathop  \to \limits^{{N_t} \to \infty } \frac{1}{{{N_t}}}\tau {N_0}\left( {\sum\limits_{p = 1}^K {\left[ {{\bf{A}}_{lt,{{5}}}^l} \right]_{pp}^2{\rm{tr}}\left( {{\bf{R}}_{lk}^0{\bf{R}}_{lp}^l} \right){\rm{tr}}\left( {{\bf{R}}_{lp}^l} \right) + \sum\limits_{p = 1}^K {\sum\limits_{m = 1,m \ne p}^K {{{\left[ {{\bf{A}}_{lt\,{\rm{5}}}^l} \right]}_{pp}}{{\left[ {{\bf{A}}_{lt\,{\rm{5}}}^l} \right]}_{mm}}{\rm{tr}}\left( {{\bf{R}}_{lp}^l{\bf{R}}_{lk}^0{\bf{R}}_{lm}^l} \right)} } } } \right).
\label{eq:hlk_v_eq_12}
\end{align}
\vspace*{6pt}

 \hrulefill
\vspace*{4pt}
\end{figure*}

Substituting (\ref{eq:hlk_v_eq_11}) and (\ref{eq:hlk_v_eq_12}) into (\ref{eq:hlk_v_eq_2}), we obtain
 \begin{align}
 &  \frac{1}{{{N_t}}}{\left( {{\bf{h}}_{lk}^0} \right)^H}{\left( {{\bf{V}}_{eq}^l} \right)^H}{{{\bf{\hat h}}}_{eq,lt}}{\bf{\hat h}}{_{eq,lt}^H}{\bf{V}}_{eq}^l{\bf{h}}_{lk}^0 \nonumber \\
& \mathop  \to \limits^{{N_t} \to \infty } \frac{1}{{{N_t}}}P_I^2{\tau ^2}{{\rm tr}^2}\left( {{\bf{R}}_{lt}^l} \right)\left[ {{\bf{A}}_{lt,{{5}}}^l} \right]_{tt}^2 {\rm tr}\left( {{\bf{R}}_{lt}^l{\bf{R}}_{lk}^0} \right) \nonumber \\
& {{ + }} \frac{1}{{{N_t}}}{P_I}\tau {N_0}\left( {\sum\limits_{p = 1}^K {\left[ {{\bf{A}}_{lt\,{{5}}}^l} \right]_{pp}^2 {\rm tr}\left( {{\bf{R}}_{lk}^0{\bf{R}}_{lp}^l} \right){\rm tr}\left( {{\bf{R}}_{lp}^l} \right) } } \right.  \nonumber \\
&  \left.  + \sum\limits_{p = 1}^K \sum\limits_{m = 1,m \ne p}^K {{\left[ {{\bf{A}}_{lt,{{5}}}^l} \right]}_{pp}}{{\left[ {{\bf{A}}_{lt,{{5}}}^l} \right]}_{mm}} {\rm tr} \left( {{\bf{R}}_{lp}^l{\bf{R}}_{lk}^0{\bf{R}}_{lm}^l} \right) \right) \nonumber \\
&  = c_{lt,1}^l. \label{eq:hlk_v_eq_3}
\end{align}

By combining (\ref{eq:gltk}), (\ref{eq:alt}), and (\ref{eq:hlk_v_eq_3}), we obtain
\begin{align}
\frac{P}{{{N_t}}}{\left| {g_{lt,k}^0} \right|^2}\mathop  \to \limits^{{N_t} \to \infty } \frac{P}{{{N_t}}}\frac{{c_{lt{\rm{,1}}}^l}}{{a_{lt{\rm{,1}}}^l}}. \label{eq:gltk_2}
\end{align}

For $C_{k}^{\rm eve}$ in (\ref{eq:C_eve_k}), we know from (\ref{eq:UY})
that when $N_t \rightarrow \infty$, $ \mathbf{V}_{eq}^{0} \mathbf{H}_e^0 \rightarrow 0$.
Therefore, we have
\begin{align} \label{c_kupper}
C_{k}^{\rm eve} \mathop \to \limits^{{N_t} \to \infty} 0 .
\end{align}

Substituting (\ref{eq:g0kk2}),  (\ref{eq:g0tk}), (\ref{eq:gltk}), and (\ref{c_kupper})
into (\ref{eq:sum_rate}) completes the proof.



\end{document}